\documentclass[pre,floatfix,twocolumn,showpacs]{revtex4}
\usepackage{amsmath}
\usepackage{amsfonts}
\usepackage{mathrsfs}
\usepackage{epsfig}
\usepackage{graphicx}
\usepackage{dcolumn}

\newcommand{\br}[1]{[#1]}
\newcommand{\dt}[1]{\frac{d #1}{dt}}

\makeatletter
\newcommand{\FF}{\afterassignment\FF@aux\count0=}
\newcommand{\FF@aux}{\csname FF\the\count0\endcsname}
\makeatother

\makeatletter
\newcommand{\F}{\afterassignment\F@aux\count0=}
\newcommand{\F@aux}{\csname F\the\count0\endcsname}
\makeatother

\makeatletter
\newcommand{\GG}{\afterassignment\GG@aux\count0=}
\newcommand{\GG@aux}{\csname GG\the\count0\endcsname}
\makeatother

\makeatletter
\newcommand{\G}{\afterassignment\G@aux\count0=}
\newcommand{\G@aux}{\csname G\the\count0\endcsname}
\makeatother

\makeatletter
\newcommand{\LL}{\afterassignment\LL@aux\count0=}
\newcommand{\LL@aux}{\csname LL\the\count0\endcsname}
\makeatother

\makeatletter
\newcommand{\LM}{\afterassignment\LM@aux\count0=}
\newcommand{\LM@aux}{\csname LM\the\count0\endcsname}
\makeatother

\makeatletter
\newcommand{\FG}{\afterassignment\FG@aux\count0=}
\newcommand{\FG@aux}{\csname FG\the\count0\endcsname}
\makeatother

\makeatletter
\newcommand{\LD}{\afterassignment\LD@aux\count0=}
\newcommand{\LD@aux}{\csname LD\the\count0\endcsname}
\makeatother

\expandafter\newcommand\csname FF1\endcsname{[3K]}
\expandafter\newcommand\csname FF2\endcsname{[S.3K]}
\expandafter\newcommand\csname FF3\endcsname{[3K^{*}.P_{3K}]}
\expandafter\newcommand\csname FF4\endcsname{[3K^{*}]}
\expandafter\newcommand\csname FF5\endcsname{[2K]}
\expandafter\newcommand\csname FF6\endcsname{[3K^{*}.2K]}
\expandafter\newcommand\csname FF7\endcsname{[2K^{*}.P_{2K}]}
\expandafter\newcommand\csname FF8\endcsname{[2K^{*}]}
\expandafter\newcommand\csname FF9\endcsname{[3K^{*}.2K^{*}]}
\expandafter\newcommand\csname FF10\endcsname{[2K^{**}.P_{2K}]}
\expandafter\newcommand\csname FF11\endcsname{[2K^{**}]}
\expandafter\newcommand\csname FF12\endcsname{[K]}
\expandafter\newcommand\csname FF13\endcsname{[2K^{**}.K]}
\expandafter\newcommand\csname FF14\endcsname{[K^{*}.P_{K}]}
\expandafter\newcommand\csname FF15\endcsname{[K^{*}]}
\expandafter\newcommand\csname FF16\endcsname{[2K^{**}.K^{*}]}
\expandafter\newcommand\csname FF17\endcsname{[K^{**}.P_{K}]}
\expandafter\newcommand\csname FF18\endcsname{[K^{**}]}

\expandafter\newcommand\csname GG1\endcsname{[P^{f}_{3K}]}
\expandafter\newcommand\csname GG2\endcsname{[P^{f}_{2K}]}
\expandafter\newcommand\csname GG3\endcsname{[P^{f}_{2K}]}
\expandafter\newcommand\csname GG4\endcsname{[P^{f}_{K}]}
\expandafter\newcommand\csname GG5\endcsname{[P^{f}_{K}]}

\expandafter\newcommand\csname F1\endcsname{F_{1}}
\expandafter\newcommand\csname F2\endcsname{F_{2}}
\expandafter\newcommand\csname F3\endcsname{F_{3}}
\expandafter\newcommand\csname F4\endcsname{F_{4}}
\expandafter\newcommand\csname F5\endcsname{F_{5}}
\expandafter\newcommand\csname F6\endcsname{F_{6}}
\expandafter\newcommand\csname F7\endcsname{F_{7}}
\expandafter\newcommand\csname F8\endcsname{F_{8}}
\expandafter\newcommand\csname F9\endcsname{F_{9}}
\expandafter\newcommand\csname F10\endcsname{F_{10}}
\expandafter\newcommand\csname F11\endcsname{F_{11}}
\expandafter\newcommand\csname F12\endcsname{F_{12}}
\expandafter\newcommand\csname F13\endcsname{F_{13}}
\expandafter\newcommand\csname F14\endcsname{F_{14}}
\expandafter\newcommand\csname F15\endcsname{F_{15}}
\expandafter\newcommand\csname F16\endcsname{F_{16}}
\expandafter\newcommand\csname F17\endcsname{F_{17}}
\expandafter\newcommand\csname F18\endcsname{F_{18}}

\expandafter\newcommand\csname G1\endcsname{G_{1}}
\expandafter\newcommand\csname G2\endcsname{G_{2}}
\expandafter\newcommand\csname G3\endcsname{G_{3}}
\expandafter\newcommand\csname G4\endcsname{G_{4}}
\expandafter\newcommand\csname G5\endcsname{G_{5}}

\expandafter\newcommand\csname LL1\endcsname{k_{-1}}
\expandafter\newcommand\csname LL2\endcsname{kp_{2}}
\expandafter\newcommand\csname LL3\endcsname{k_{1}}
\expandafter\newcommand\csname LL4\endcsname{k_{2}}
\expandafter\newcommand\csname LL5\endcsname{kp_{1}}
\expandafter\newcommand\csname LL6\endcsname{kp_{-1}}
\expandafter\newcommand\csname LL7\endcsname{k_{-3}}
\expandafter\newcommand\csname LL8\endcsname{k_{4}}
\expandafter\newcommand\csname LL9\endcsname{k_{3}}
\expandafter\newcommand\csname LL10\endcsname{k_{-5}}
\expandafter\newcommand\csname LL11\endcsname{k_{6}}
\expandafter\newcommand\csname LL12\endcsname{k_{5}}
\expandafter\newcommand\csname LL19\endcsname{kp_{4}}
\expandafter\newcommand\csname LL20\endcsname{kp_{3}}
\expandafter\newcommand\csname LL21\endcsname{kp_{-3}}
\expandafter\newcommand\csname LL22\endcsname{kp_{5}}
\expandafter\newcommand\csname LL23\endcsname{kp_{6}}
\expandafter\newcommand\csname LL24\endcsname{kp_{-5}}
\expandafter\newcommand\csname LL25\endcsname{k_{-7}}
\expandafter\newcommand\csname LL26\endcsname{k_{8}}
\expandafter\newcommand\csname LL27\endcsname{k_{7}}
\expandafter\newcommand\csname LL28\endcsname{k_{-9}}
\expandafter\newcommand\csname LL29\endcsname{k_{10}}
\expandafter\newcommand\csname LL30\endcsname{k_{9}}
\expandafter\newcommand\csname LL31\endcsname{kp_{8}}
\expandafter\newcommand\csname LL32\endcsname{kp_{7}}
\expandafter\newcommand\csname LL33\endcsname{kp_{-7}}
\expandafter\newcommand\csname LL34\endcsname{kp_{10}}
\expandafter\newcommand\csname LL35\endcsname{kp_{9}}
\expandafter\newcommand\csname LL36\endcsname{kp_{-9}}

\expandafter\newcommand\csname LM1\endcsname{\lambda_{1}}
\expandafter\newcommand\csname LM2\endcsname{\lambda_{2}}
\expandafter\newcommand\csname LM3\endcsname{\lambda_{3}}
\expandafter\newcommand\csname LM4\endcsname{\lambda_{4}}
\expandafter\newcommand\csname LM5\endcsname{\lambda_{5}}
\expandafter\newcommand\csname LM6\endcsname{\lambda_{6}}
\expandafter\newcommand\csname LM7\endcsname{\lambda_{7}}
\expandafter\newcommand\csname LM8\endcsname{\lambda_{8}}
\expandafter\newcommand\csname LM9\endcsname{\lambda_{9}}
\expandafter\newcommand\csname LM10\endcsname{\lambda_{10}}
\expandafter\newcommand\csname LM11\endcsname{\lambda_{11}}
\expandafter\newcommand\csname LM12\endcsname{\lambda_{12}}
\expandafter\newcommand\csname LM13\endcsname{\lambda_{13}}
\expandafter\newcommand\csname LM14\endcsname{\lambda_{14}}
\expandafter\newcommand\csname LM15\endcsname{\lambda_{15}}
\expandafter\newcommand\csname LM16\endcsname{\lambda_{16}}
\expandafter\newcommand\csname LM17\endcsname{\lambda_{17}}
\expandafter\newcommand\csname LM18\endcsname{\lambda_{18}}
\expandafter\newcommand\csname LM19\endcsname{\lambda_{19}}
\expandafter\newcommand\csname LM20\endcsname{\lambda_{20}}
\expandafter\newcommand\csname LM21\endcsname{\lambda_{21}}
\expandafter\newcommand\csname LM22\endcsname{\lambda_{22}}
\expandafter\newcommand\csname LM23\endcsname{\lambda_{23}}
\expandafter\newcommand\csname LM24\endcsname{\lambda_{24}}
\expandafter\newcommand\csname LM25\endcsname{\lambda_{25}}
\expandafter\newcommand\csname LM26\endcsname{\lambda_{26}}
\expandafter\newcommand\csname LM27\endcsname{\lambda_{27}}
\expandafter\newcommand\csname LM28\endcsname{\lambda_{28}}
\expandafter\newcommand\csname LM29\endcsname{\lambda_{29}}
\expandafter\newcommand\csname LM30\endcsname{\lambda_{30}}
\expandafter\newcommand\csname LM31\endcsname{\lambda_{31}}
\expandafter\newcommand\csname LM32\endcsname{\lambda_{32}}
\expandafter\newcommand\csname LM33\endcsname{\lambda_{33}}
\expandafter\newcommand\csname LM34\endcsname{\lambda_{34}}
\expandafter\newcommand\csname LM35\endcsname{\lambda_{35}}
\expandafter\newcommand\csname LM36\endcsname{\lambda_{36}}
\expandafter\newcommand\csname LM37\endcsname{\lambda_{37}}
\expandafter\newcommand\csname LM38\endcsname{\lambda_{38}}
\expandafter\newcommand\csname LM39\endcsname{\lambda_{39}}
\expandafter\newcommand\csname LM40\endcsname{\lambda_{40}}
\expandafter\newcommand\csname LM41\endcsname{\lambda_{41}}
\expandafter\newcommand\csname LM42\endcsname{\lambda_{42}}
\expandafter\newcommand\csname LM43\endcsname{\lambda_{43}}
\expandafter\newcommand\csname LM44\endcsname{\lambda_{44}}
\expandafter\newcommand\csname LM45\endcsname{\lambda_{45}}
\expandafter\newcommand\csname LM46\endcsname{\lambda_{46}}
\expandafter\newcommand\csname LM47\endcsname{\lambda_{47}}
\expandafter\newcommand\csname LM48\endcsname{\lambda_{48}}
\expandafter\newcommand\csname LM49\endcsname{\lambda_{49}}
\expandafter\newcommand\csname LM50\endcsname{\lambda_{50}}
\expandafter\newcommand\csname LM51\endcsname{\lambda_{51}}
\expandafter\newcommand\csname LM52\endcsname{\lambda_{52}}
\expandafter\newcommand\csname LM53\endcsname{\lambda_{53}}
\expandafter\newcommand\csname LM54\endcsname{\lambda_{54}}

\expandafter\newcommand\csname FG1\endcsname{\gamma_{1}\,F_{1}^{\dagger}}
\expandafter\newcommand\csname FG2\endcsname{\gamma_{2}\,F_{2}^{\dagger}}
\expandafter\newcommand\csname FG3\endcsname{\gamma_{3}\,F_{3}^{\dagger}}
\expandafter\newcommand\csname FG4\endcsname{\gamma_{4}\,F_{4}^{\dagger}}
\expandafter\newcommand\csname FG5\endcsname{\gamma_{5}\,F_{5}^{\dagger}}
\expandafter\newcommand\csname FG6\endcsname{\gamma_{6}\,F_{6}^{\dagger}}
\expandafter\newcommand\csname FG7\endcsname{\gamma_{7}\,F_{7}^{\dagger}}
\expandafter\newcommand\csname FG8\endcsname{\gamma_{8}\,F_{8}^{\dagger}}
\expandafter\newcommand\csname FG9\endcsname{\gamma_{9}\,F_{9}^{\dagger}}
\expandafter\newcommand\csname FG10\endcsname{\gamma_{10}\,F_{10}^{\dagger}}
\expandafter\newcommand\csname FG11\endcsname{\gamma_{11}\,F_{11}^{\dagger}}
\expandafter\newcommand\csname FG12\endcsname{\gamma_{12}\,F_{12}^{\dagger}}
\expandafter\newcommand\csname FG13\endcsname{\gamma_{13}\,F_{13}^{\dagger}}
\expandafter\newcommand\csname FG14\endcsname{\gamma_{14}\,F_{14}^{\dagger}}
\expandafter\newcommand\csname FG15\endcsname{\gamma_{15}\,F_{15}^{\dagger}}
\expandafter\newcommand\csname FG16\endcsname{\gamma_{16}\,F_{16}^{\dagger}}
\expandafter\newcommand\csname FG17\endcsname{\gamma_{17}\,F_{17}^{\dagger}}
\expandafter\newcommand\csname FG18\endcsname{\gamma_{18}\,F_{18}^{\dagger}}

\begin{document}
\title{Emergent memory in cell signaling: 
Persistent adaptive dynamics in cascades can arise from the diversity of
relaxation time-scales}

\author{Tanmay Mitra$^{1,2}$, Shakti N. Menon$^1$ and Sitabhra
Sinha$^{1,2}$}
\affiliation{$^1$The Institute of Mathematical Sciences, CIT Campus,
Taramani, Chennai 600113, India.\\
$^2$Homi Bhabha National Institute, Anushaktinagar, Mumbai 400
094, India.}
\date{\today}
\begin{abstract}
The mitogen-activated protein kinase (MAPK) signaling cascade, an 
evolutionarily conserved motif present in all eukaryotic cells, is 
involved in coordinating critical cell-fate decisions, regulating protein
synthesis, and mediating learning and memory. 
While the steady-state behavior of the pathway stimulated by a
time-invariant signal is relatively well-understood, 
we show using a computational model 
that it exhibits a rich repertoire of transient adaptive
responses to changes in stimuli.
When the signal is switched on, the response is characterized by
long-lived modulations in
frequency as well as amplitude.
On withdrawing the stimulus, the activity decays over timescales much
longer than that of phosphorylation-dephosphorylation processes, exhibiting
reverberations characterized by repeated spiking in the activated MAPK
concentration.
The long-term persistence of such post-stimulus activity suggests that the
cascade retains memory of the signal for a significant duration following its
removal, even in the absence of any explicit feedback or cross-talk
with other pathways.
We find that the molecular mechanism underlying this behavior 
is related to the existence of distinct relaxation rates for the
different cascade components. 
This results in 
the imbalance
of fluxes between different layers of the cascade, with the repeated 
reuse of
activated kinases as enzymes when they are released from
sequestration in complexes leading to one or more spike events
following the removal of the stimulus.
The persistent adaptive response reported here, indicative of a cellular
``short-term'' memory, suggests that
this ubiquitous signaling pathway plays an even more central role
in information processing by eukaryotic cells.
\end{abstract}

\maketitle
\newpage

\section{Introduction}
Intra-cellular signaling networks are paradigmatic examples
of complex adaptive systems that exhibit a rich
repertoire of responses to stimuli~\cite{IyengarEm1999}.
Such networks mediate the response of a cell to a wide variety of
extra- and intra-cellular signals primarily through a sequence of
enzyme-substrate biochemical reactions~\cite{Lahav2013,Kholodenko2006}.
While the complexity of the entire signaling system is daunting~\cite{Iyengar1999},
it is possible to gain an insight into how it functions by focusing on
a key set of frequently occurring motifs. These
often take the form of linear signaling cascades, referred to as
pathways.
One of the best known of these pathways is the mitogen-activated protein
kinase (MAPK) cascade that is present in all eukaryotic
cells~\cite{Johnson1999,Seger1995}. It is
involved in regulating a range of vital cellular functions, including 
proliferation and apoptosis~\cite{Seger1995}, stress
response~\cite{Cooper1995} and gene expression~\cite{Karin1996}.
This signaling module comprises a sequential
arrangement of three protein kinases, viz., MAPK, MAPK kinase (MAP2K)
and MAPK kinase kinase (MAP3K).
Modular function is initiated when            
extracellular signals stimulate membrane-bound receptors upstream of
the cascade, with the
information being relayed to MAP3K by a series of intermediaries.
Activated kinases in each layer of the module function as enzymes for
phosphorylating (and thereby activating) the kinase in the level
immediately downstream, with the subsequent deactivation being mediated by
corresponding dephosphorylating enzymes known as phosphatases
(P'ase).
The terminal kinase in this cascade, i.e., MAPK, transmits the signal
further downstream by phosphorylating various proteins including
transcription regulators~\cite{Alberts6thed}.
Extensive investigations into the steady-state behavior of the cascade
have contributed towards an in-depth understanding of several emergent
features including ultrasensitivity~\cite{Ferrell1996}, and 
oscillations~\cite{Shankaran2009,Qiao2007} that arise through
retrograde propagation of activity~\cite{Sitabhra2013,Ventura2008}.
One of the striking features of the cascade is the occurrence of
bistability which allows the system to switch between two
possible states corresponding to low and high activity~\cite{Kholodenko2004,Herzel2007,KholodenkoBS2006,Qiao2007,Ferrell2002}. This provides
a post-transcriptional mechanism for obtaining a sustained response from
transient signals, i.e., cellular memory~\cite{Inniss2013,Xiong2003}. 

Memory can be understood as long-term alterations in the state of a
system in response to environmental changes, which allow the system to retain
information about transient signals long after being exposed to 
them~\cite{Inniss2013}.
This can arise in the cell through mechanisms such as auto-regulatory
transcriptional positive feedback~\cite{Silver2007} and nucleosomal
modifications~\cite{Burrill2010}. In the context of cell-fate
determination,
it has been shown that an irreversible biochemical response can be
generated from a short-lived stimulus through feedback-based
bistability~\cite{Xiong2003}. This corresponds to a permanent
alteration of the state
of the system, thereby actively maintaining `memory' of the
signal. As bistability has also been observed to arise through
multi-site phosphorylation in signaling modules, 
protein phosphorylation has been suggested
as a plausible post-transcriptional mechanism for cellular
memory~\cite{Gunawardena2005,Hadac2013,Inniss2013}. 
In particular, there have been extensive investigations of the
MAPK cascade
as it integrates a large range of signals received by the
cell in order to control numerous cellular decisions~\cite{Carew2013,Tsien2001,Sweatt2001,Schaller2004,Bonnet2005,Flavell2002}.
While these investigations have considered the steady state behavior
of the system, one may also observe
transitory modulations in the response of the cascade in a changing
environment. The latter could encode
information about prior stimuli to which the system was exposed, and
can be a
potential mechanism for imparting a form of ``short-term'' memory to the
signaling cascade.

In this paper we show that a linear MAPK cascade can indeed exhibit
short-term memory through transient modulations in its response to an environmental change. 
Crucially, this can arise even in the absence of explicit feedback
between different layers or cross-talk with other pathways.
These modulations can persist long after the initial trigger, lasting
for durations that are several orders of magnitude longer than the
time-scales associated with phosphorylation-dephosphorylation
processes. We demonstrate that this occurs both when a signal begins activating
the MAPK cascade, as well as when it is withdrawn. On application of
the stimulus, the module exhibits long-lived frequency and amplitude
modulations in the activation profile of the constituent kinases. 
Following the withdrawal of stimulus, activity in the cascade decays
over an extremely long
time-scale, during which reverberatory dynamics, characterized by
large-amplitude spiking in MAP Kinase activity, can be observed.
We explain the emergence of such long-lived memory of the withdrawn
stimulus in terms of the
imbalance of fluxes between different layers of the cascade, which
results from the diversity of relaxation time-scales of the cascade
components, and the reuse
of activated kinases as enzymes when they are released from
sequestration.
This phenomenon is seen to be robust with respect to
variations in the molecular concentrations of the constituent kinases
and phosphatases. Our results reveal that a biochemical
signaling module as simple as the MAPK cascade is capable of
exhibiting short-term memory that is manifested as persistent
modulations in the adaptive response of the system to changes in stimuli.


\section{Methods}
The dynamics of the three layer MAPK signaling cascade has
been simulated using the Huang-Ferrell model~\cite{Ferrell1996}.
Each of the constituent kinase and phosphatase-mediated
enzyme-substrate reactions comprise (i) a reversible
step corresponding to the formation of the enzyme-substrate complex
and (ii) an irreversible product formation step corresponding to the
activation/deactivation of a kinase, as described in the Supplementary
Information.
The time-evolution of the molecular concentrations of the different components
of the cascade are modeled using a set of
coupled ordinary differential equations (see Supplementary
Information) that are integrated using the stiff solver
\texttt{ode15s} implemented in {\em MATLAB Release 2010b}. 
Note that the quasi-steady-state hypothesis has not been
invoked~\cite{Wolkenhauer2007}.
To ensure that initially all kinases are non-phosphorylated we prepare
the initial resting state of the system by simulating it for a long duration
($\sim 10^6$ mins) in the absence of any signal.
Subsequently MAP3K is exposed to a stimulus of amplitude $S$ and
duration $5000$ minutes. Following the removal of the stimulus, we
continue to simulate the system until it returns to the resting state
or the simulation duration exceeds $10^4$ minutes.

We have analyzed the long-lived reverberatory activity of the cascade
after the removal of the stimulus by using the following measures:

\noindent
(i) {\em The primary recovery time} ($\tau_{\rm PR}$). Following the
activation of the cascade by introducing a stimulus, the maximum
concentration $R_{\rm max}$ of MAPK$^{**}$ is recorded (Note that
$^{**}$ represents a doubly phosphorylated kinase while $^*$
indicates that it is singly
phosphorylated). 
On removing the stimulus, MAPK activity starts to decay. The time
taken for MAPK$^{**}$ to monotonically decrease to half of $R_{\rm
max}$ is defined as the primary recovery time ($\tau_{\rm PR}$).

\noindent
(ii) {\em Number of spikes during relaxation} ($N_r$). 
Following primary recovery, MAPK activity may exhibit
a series of spikes, which are defined to be occurring whenever
MAPK$^{**}$ concentration
exceeds $70\%$ of $R_{\rm max}$.
The number of such spikes that are observed before the cascade reaches
its resting state is designated as $N_r$.

\noindent
(iii) {\em The total duration of reverberatory activity} ($\tau_r$).
When spiking is observed in MAPK activity following the removal of the
applied stimulus, the reverberatory activity duration is defined as
the interval between the termination of primary recovery and the final
spike event, i.e., $\tau_{r} = t_{final} - \tau_{PR}$. The time of the
$i$th spike $t_i$ is defined as the instant when MAPK activity reaches
maximum during that particular event.
For $\tau_{PR} > 6000$ mins, the total duration of the reverberatory
activity may not be measured accurately as the total simulation
duration does not exceed $10^4$ minutes. 

\noindent
(iv) {\em The total memory time} ($\tau_m$). The total duration of
memory activity 
following removal of the applied stimulus is defined as the sum of the
primary recovery time and the total duration of reverberatory
activity, i.e., $\tau_{m} = \tau_{PR} + \tau_{r}$.
Note that when the steady-state behavior of the cascade in presence of
the signal is oscillatory, on withdrawing the signal the activity may
decay extremely rapidly resulting in $\tau_{m} \approx 0$.

\noindent
(v) {\em Relaxation time} ($\tau_{\rm x}$). 
For the situations where the steady state corresponds to a
fixed-point attractor we define a relaxation time $\tau_{\rm x}$ 
for each constituent of the cascade. This is 
the time required by its concentration
to evolve to the half-way point between the resting state and
steady state values.

\begin{figure} 
\begin{center}
\includegraphics[width=0.95\linewidth]{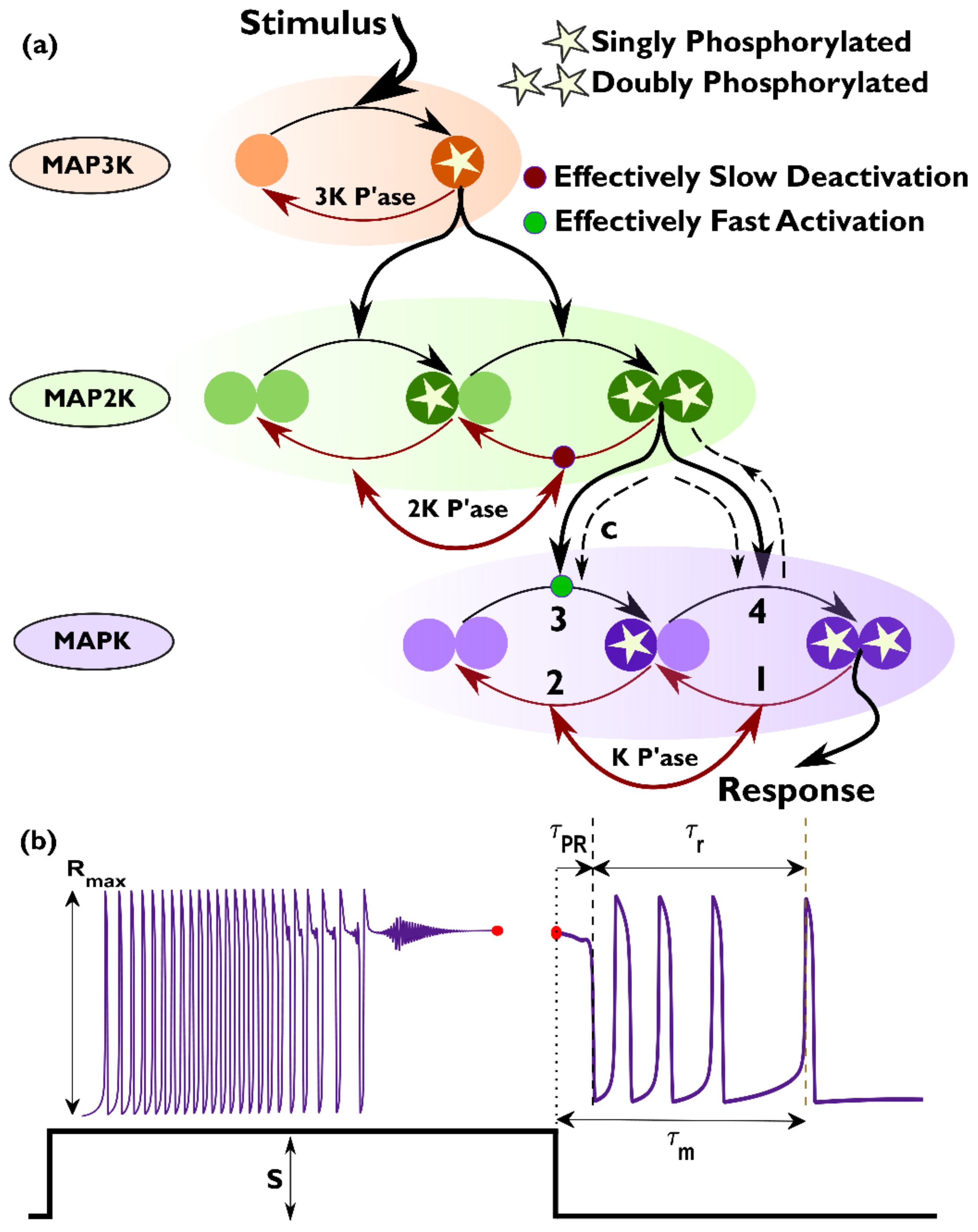}
\end{center}
\caption{Adaptive response of MAPK cascade to a changing stimulus.
(a) Schematic representation of a linear MAPK cascade comprising
three layers. Signaling is initiated by a stimulus S activating
MAPK kinase kinase (MAP3K). Activation/deactivation of kinases is
achieved by adding/removing phosphate groups, which is referred to as
phosphorylation/dephosphorylation respectively. The activated MAP3K
regulates the phosphorylation of MAPK kinase (MAP2K).
Doubly phosphorylated MAP2K, in its turn, controls the activation of
MAPK. The response of the cascade to the signal is measured in terms
of MAPK activity, viz., the concentration of doubly phosphorylated MAPK. 
Deactivation of a phosphorylated kinase is regulated by the
corresponding phosphatase (indicated by P'ase) in the corresponding
layer of the cascade. 
The numbers $1 - 4$ represent the sequence of events that lead to the
emergence of a large amplitude spiking response following
the withdrawal of the stimulus. The enzyme-substrate protein complex
formed during activation of MAPK by doubly phosphorylated MAP2K is
indicated by ``c''. Broken lines have been used to highlight the principal
processes that drive the reverberatory
dynamics, which functions as a memory of the signal (see text for details).
(b) Schematic illustrating the emergence of long-lived transient
modulations of MAPK activity in response to initiation of a signal of
optimal strength $S$. 
Withdrawing the stimulus can result in persistent large-amplitude
spiking in the response of MAPK, suggestive of a form of
``short-term'' memory. The maximum response of MAPK to the stimulus is
denoted by $R_{\rm max}$. The primary recovery time ($\tau_{PR}$) is
characterized as the duration following withdrawal of stimulus after
which MAPK activity decreases to its half-maximum value ($R_{\rm
max}/2$) for the first time. 
The duration over which reverberatory dynamics occurs is
indicated by $\tau_{r}$, while the total duration for which memory of
the withdrawn stimulus persists is $\tau_{m} = \tau_{PR} + \tau_{r}$. 
}
\label{fig:fig1}
\end{figure}


\section{Results}
{\em Emergence of persistent modulations in kinase activity.}
For the results reported in this paper we consider the Huang-Ferrell
model of the MAPK signaling cascade~\cite{Ferrell1996},
schematically shown in Fig.~\ref{fig:fig1}~(a). Typically, investigations
into the dynamics of this model focus on the steady-state response to
sustained stimulation. In contrast, here we report on the transient
activity of the system following a change in the stimulus.
Specifically, we describe the response immediately following the
introduction of a
signal of amplitude $S$ and that following its removal.
Our results reveal that such transients can be unexpectedly
long-lived, lasting for durations that are much longer compared to
the time-scales associated with the phosphorylation and dephosphorylation
processes in the cascade (Fig.~\ref{fig:fig1},~b).

\begin{figure} [hbt!] 
\begin{center}
\includegraphics[ ]{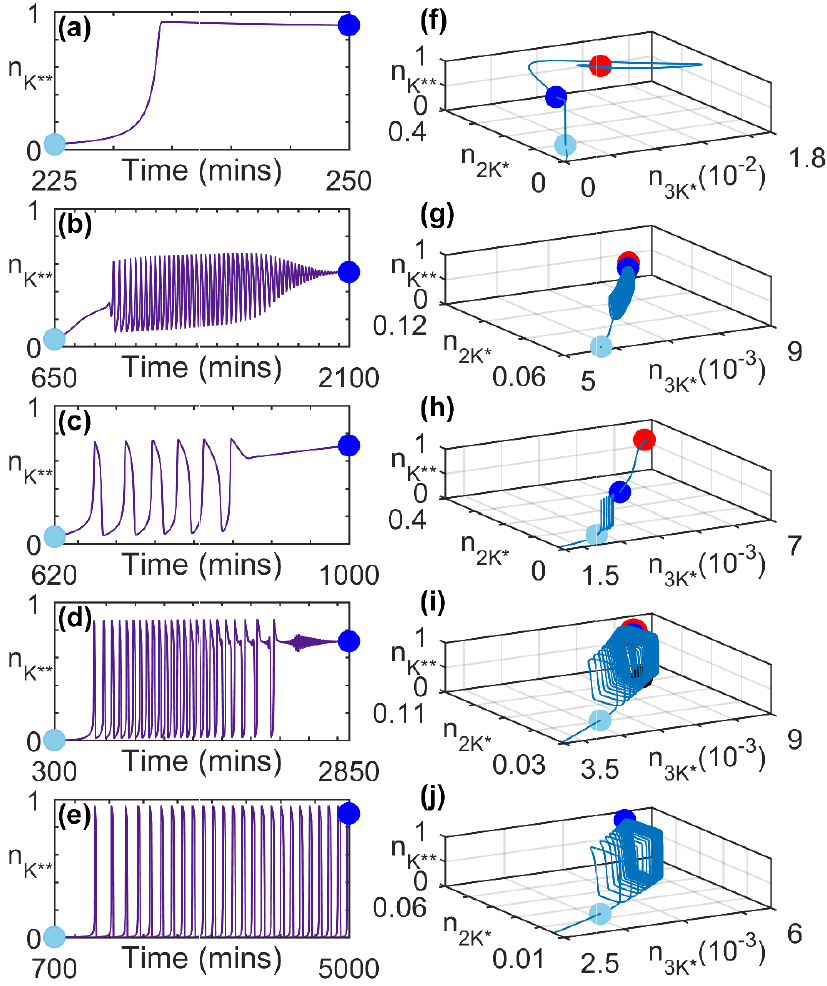}
\end{center}
\caption{Transient activity in MAPK cascade immediately following the
application of a stimulus having amplitude $S =
1.2 \times 10^{-6} \mu M$ at $t=0$. (a-e) Characteristic time
series for the normalized concentration of doubly phosphorylated MAPK
($n_{K^{**}}$) shown
for different total concentrations of kinases. 
The concentration of active MAPK is
insignificant prior to the time periods shown in panels (a-e).
(f-j) Trajectories representing the evolution of the systems in panels (a-e)
in the projection of the phase-space
on the planes comprising normalized concentrations of active MAP3K ($n_{3K^{*}}$),
singly phosphorylated MAP2K ($n_{2K^{*}}$) and active MAPK ($n_{K^{**}}$). The
concentrations have been normalized by the total concentration of
MAP3K ($[3K]_{tot}$), MAP2K ($[2K]_{tot}$) and MAPK ($[K]_{tot}$),
respectively. The light blue and dark blue markers in each of the
panels (f-j) 
demarcate the portion of the trajectories that correspond to the time
series shown in panels (a-e). 
The steady state of the system is represented by a red marker in
panels (f-i). In panels (e) and (j), the system converges to a stable
limit cycle. 
For details of parameter values
for the systems shown in each of the panels see Supplementary
Information.
}
\label{fig:fig2}
\end{figure}

We first report the behavior of a cascade that is initially in the
resting state (characterized by the absence of any phosphorylated
components) when it is exposed to a signal. The transient activity
that immediately follows exhibits several non-trivial features such
as regular spiking in the activity of MAP2K and MAPK depending on the
total concentrations of the kinases
(Fig.~\ref{fig:fig2}, b-e) and the signal strength.
For a fixed initial state and signal strength,
the spikes can further show modulation in their frequency
(Fig.~\ref{fig:fig2}, c-e) as well as amplitude
(Fig.~\ref{fig:fig2}, b and d). In certain cases, both types of
modulation can be observed (Fig.~\ref{fig:fig2},~d).
In the representative time series of MAPK activity shown in
Fig.~\ref{fig:fig2}(a-e), the system dynamics eventually converges
to a stable fixed point (Fig.~\ref{fig:fig2}, a-d) or a stable limit cycle
(Fig.~\ref{fig:fig2}, e), with the attractors being independent of initial
conditions (corresponding phase space projections are shown in
Fig.~\ref{fig:fig2}, f-i). Note that when phosphorylated components
are initially present, the system reaches the asymptotic state faster.

\begin{figure} [hbt!] 
\begin{center}
\includegraphics[]{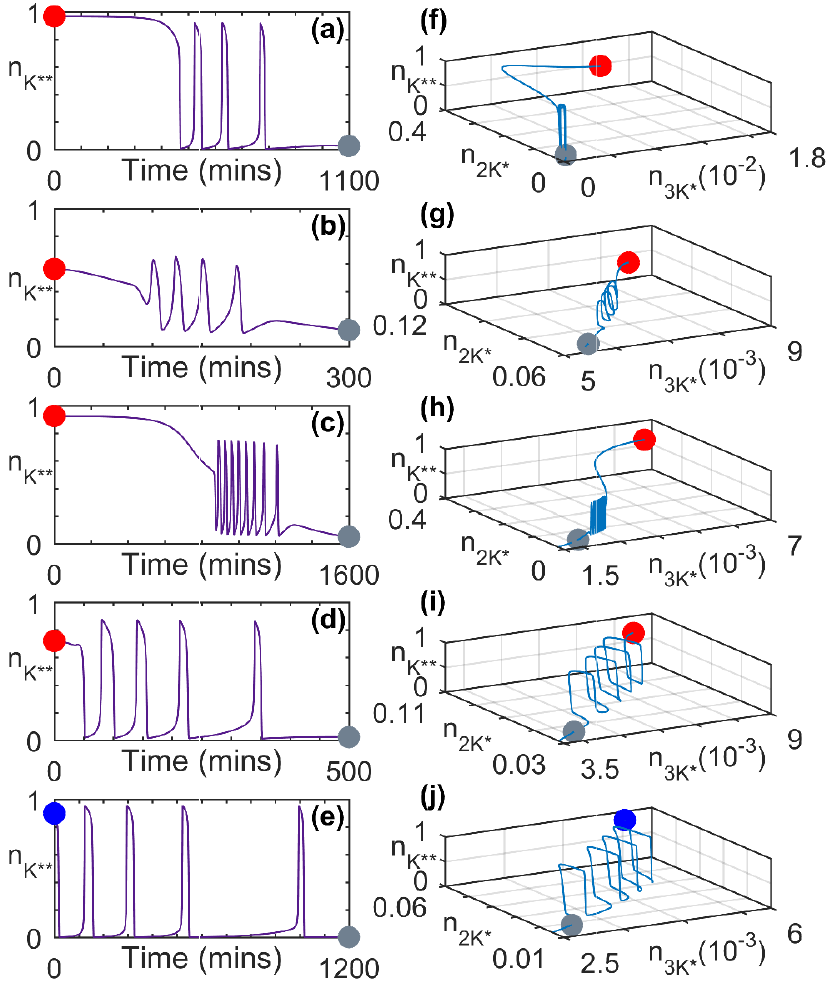}
\end{center}
\caption{Transient activity in MAPK cascade immediately following the
withdrawal (at $t=0$) of an applied stimulus having
amplitude $S = 1.2 \times 10^{-6} \mu M$. (a-e)
Characteristic time series for the normalized concentration of
doubly phosphorylated MAPK ($n_{K^{**}}$) shown for different total
concentrations of kinases. (f-j) Trajectories representing the evolution of the systems in panels (a-e) in the
projection of the phase-space on the planes comprising normalized
concentrations of active MAP3K
($n_{3K^{*}}$), singly phosphorylated MAPK ($n_{2K^{*}}$) and active MAPK
($n_{K^{**}}$). The concentrations have been normalized by the total
concentration of MAP3K ($[3K]_{tot}$), MAP2K ($[2K]_{tot}$) and MAPK
($[K]_{tot}$), respectively. The steady state of the system prior to
the withdrawal of the stimulus is represented by a 
red marker (panels f-i). The system in panels (e) and (j) is seen to
relax from a state characterized by stable limit cycle oscillations
(represented by the blue marker).
In each trajectory shown in (f-j) the grey marker denotes the state of
the system corresponding to the final time point in panels (a-e).
The concentration of active MAPK
is close to its resting state value
following the time period shown in (a-e).
The parameter values for each panel are same as those for the
corresponding panels in Fig.~\ref{fig:fig2}.
}
\label{fig:fig3}
\end{figure}

When the signal is withdrawn, the signaling
cascade can respond with large-amplitude spiking behavior in the MAPK
activity before eventually relaxing to the resting state
(Fig.~\ref{fig:fig3}). These phenomena are seen for a range of
stimuli strengths and are indicative of a form of memory that can be
achieved without explicit feedback or inter-pathway crosstalk. An
essential condition for observing the reverberatory activity is that 
prior to withdrawing the applied stimulus, the system state has been
driven above the low-amplitude response regime.
The complex modulations seen in these figures may arise
as a result of
coexisting attractors. For example, in Fig.~\ref{fig:fig2}~(d) 
the system state spends considerable time in the basin of attraction
of a limit cycle before approaching a stable fixed point (see Supplementary Information for
details).  

\begin{figure} [hbt!] 
\begin{center}
\includegraphics[width=0.99\linewidth]{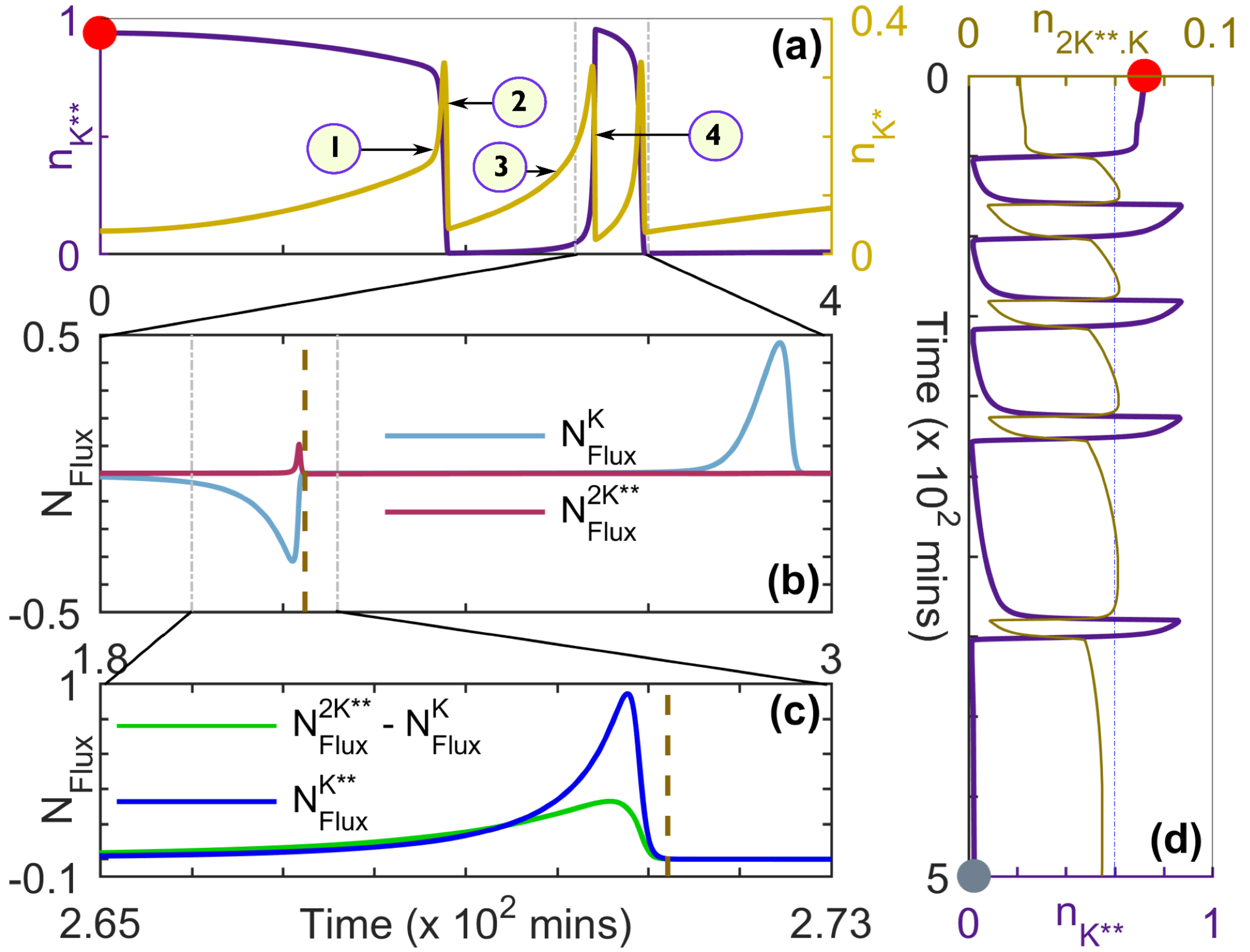}
\end{center}
\caption{Processes underlying emergent memory and reverberatory dynamics in the
MAPK cascade. 
(a) A characteristic
time-series for the normalized concentrations of singly and doubly
phosphorylated MAPK
($n_{K^{*}}$ and $n_{K^{**}}$, respectively) following the removal of
an applied stimulus of
amplitude $S = 2.0 \times 10^{-6} \mu M$ at $t = 0$. The numbers ($1 -
4$) represent the sequence of events that lead to the emergence of
the post-stimulus large-amplitude spiking activity shown schematically in
Fig.~\ref{fig:fig1}~(b). 
(b) Normalized chemical flux $N_{\rm Flux}$ of MAPK
and MAP2K$^{**}$ shown for the segment of
the time-series 
where the
spiking behavior in $n_{K^{**}}$ is observed following the withdrawal
of the stimulus to MAP3K
[demarcated by broken vertical lines in (a)].
(c) Normalized chemical flux $N_{\rm Flux}$ of
MAPK$^{**}$ shown
along with the difference between the normalized fluxes of
MAP2K$^{**}$ and MAPK for the duration indicated
by broken vertical lines in (b) corresponding to the peak in the
spiking activity of MAPK$^{**}$.
For both panels (b) and (c), normalization of flux is with respect to
the maximum of the flux for 
MAPK$^{**}$. (d) Characteristic time-series for the reverberatory
activity of MAPK following the withdrawal of a stimulus of amplitude $S
= 1.2 \times 10^{-6} \mu M$ at $t = 0$, 
showing the normalized concentration of MAPK$^{**}$
($n_{K^{**}}$) along with that of the
protein complex MAP2K$^{**}$.MAPK ($n_{2K^{**}.K}$ =
[MAP2K$^{**}$.MAPK]/$[2K]_{tot}$).
The reference line shows that the peak normalized concentration of the
protein complex eventually decreases over time.
For details of parameter values for (a-c) see Supplementary Information.
The parameter values for panel (d) are same as those
in Fig.~\ref{fig:fig3}(d).
The steady state of the system prior to the withdrawal of the stimulus
is represented by a red marker [panels (a) and (d)] while the grey
marker in (d) corresponds the final time point in 
Fig.~\ref{fig:fig3}(d).
}
\label{fig:fig4}
\end{figure}


\begin{figure} [hbt!] 
\begin{center}
\includegraphics[ ]{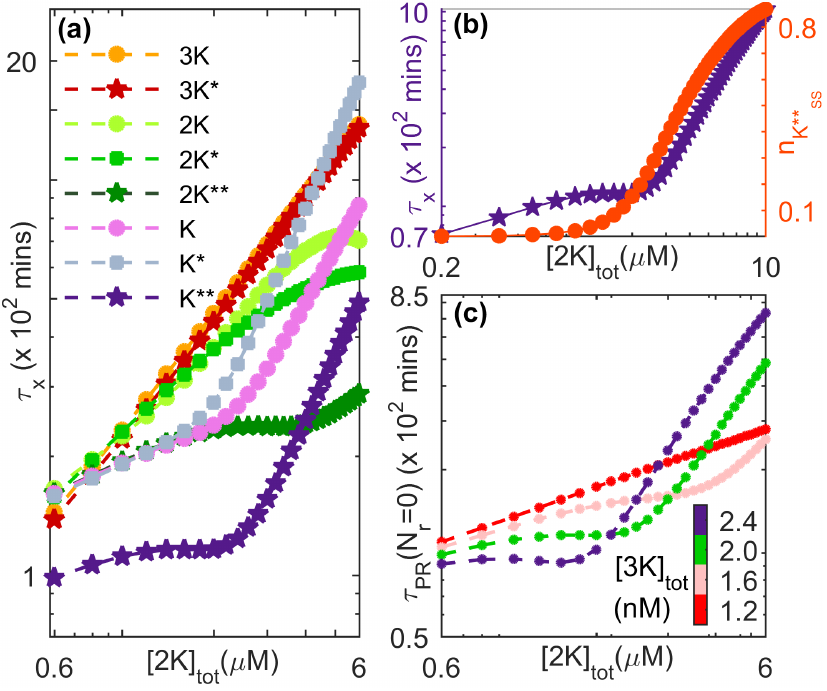}
\end{center}
\caption{Components of the MAPK cascade exhibit relaxation
behavior occurring over a broad range of time-scales.
Decay of activity is shown after withdrawing an
applied stimulus of amplitude $S = 1.2 \times 10^{-6} \mu M$. (a)
The relaxation times $\tau_{\rm x}$ of the different
molecular species (non- , singly- and doubly phosphorylated kinase
proteins) in each of the layers of the cascade vary with the
total concentration of MAP2K. The nature of this dependence is
distinct for lower and higher values of $[2K]_{tot}$, which is most
prominently observed in the lower layers of the cascade.
(b) The occurrence of distinct regimes in the relaxation behavior of
MAPK$^{**}$ for different $[2K]_{tot}$ is related to the corresponding 
increase in the steady state value attained by MAPK$^{**}$ concentration under
sustained stimulation of the cascade. At a specific value of the
steady-state normalized MAPK activity
$n_{K^{**}}$, we observe a crossover from
the regime characterized by 
slowly increasing $\tau_{\rm x}$ seen at lower total concentrations of MAP2K
to a regime where $\tau_{\rm x}$ increases relatively rapidly for higher
$[2K]_{tot}$.
(c) The crossover behavior is also observed in the dependence of the 
closely related measure $\tau_{\rm PR}$, the primary recovery time
(see Methods), on $[2K]_{tot}$. The difference between the two regimes
become more prominent upon increasing the total concentration of MAP3K
($[3K]_{tot}$). 
For both panels (a) and (b) 
$[K]_{tot} = 0.8 \mu M$ and $[3K]_{tot} = 2.0 nM$, while for panel (c),
$[K]_{tot} = 0.8  \mu M$. For details of all other parameter values see 
Supplementary Information.
}
\label{fig:fig5}
\end{figure}

\begin{figure} [hbt!] 
\begin{center}
\includegraphics[]{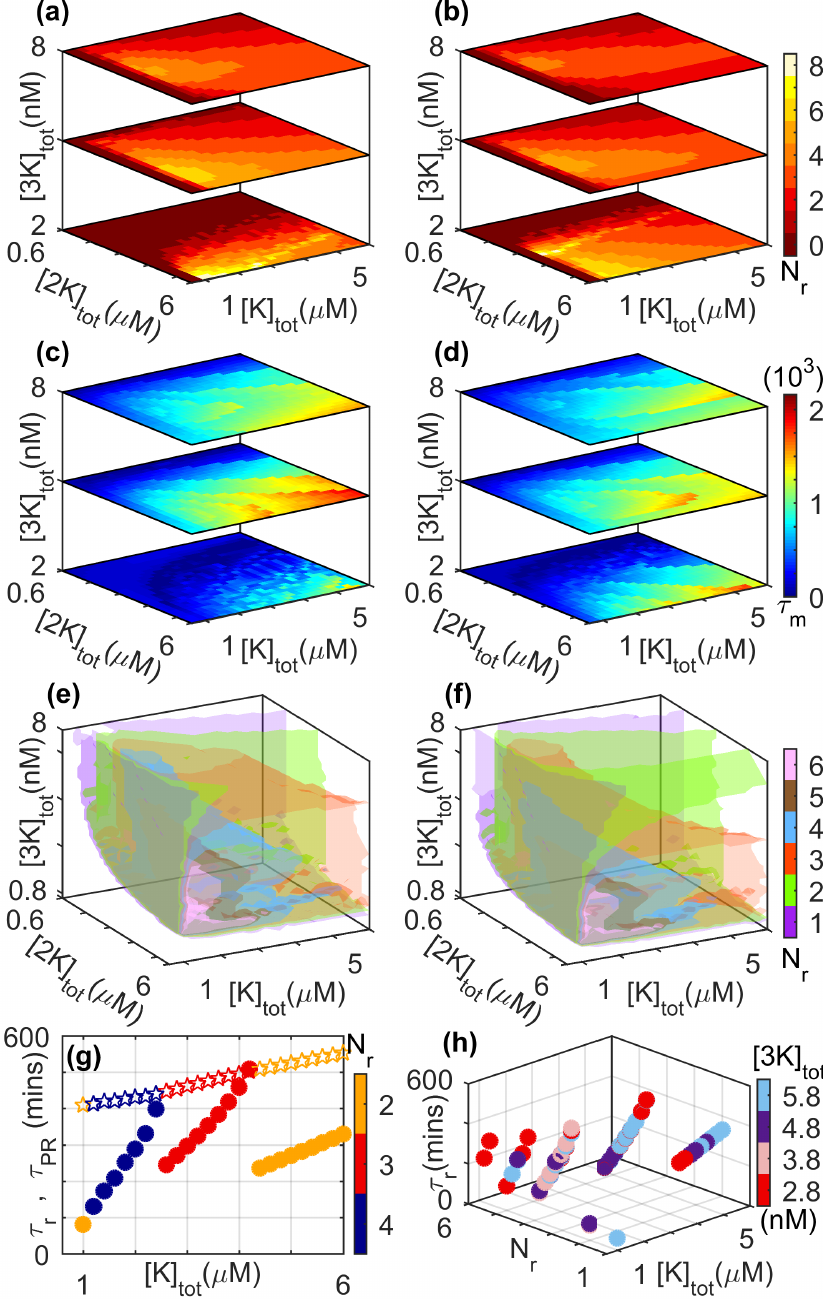}
\end{center}
\caption{Dependence of reverberatory activity on the total kinase
concentrations, viz., $[K]_{tot}$,
$[2K]_{tot}$ and $[3K]_{tot}$. 
(a-b) The number of spikes $N_{r}$, (c-d) the total
memory time $\tau_{m}$ (in minutes) and (e-f) isosurfaces for $N_r$ 
observed on withdrawing an applied stimulus
of amplitude $S$ [$=0.8 \times 10^{-6} \mu M$ for (a,c,e) and $1.2
\times 10^{-6} \mu M$ for (b,d,f)] are shown as functions of
total concentrations of the three kinases.
(g) The primary recovery time $\tau_{\rm PR}$
(stars) and the total duration of reverberatory activity $\tau_{r}$
(filled circles) are shown for different values of $N_r$ (indicated by
the color bar). While $\tau_{\rm PR}$ increases monotonically with increasing
total MAPK concentration, $\tau_{r}$ shows a more complex dependence ($[2K]_{tot} = 3 \mu M$ and $[3K]_{tot} = 4
nM$). (h) The dependence of $\tau_{r}$ on $[K]_{tot}$ for different
values of $N_{r}$ has a similar nature for different choices of
$[3K]_{tot}$ (indicated by the color bar, $[2K]_{tot} = 3 \mu M$). Note
that for panel (h), we consider only situations where the
system attains a steady state on maintaining stimulation.
For details of all other parameter values 
see Supplementary Information. 
}
\label{fig:fig6}
\end{figure}

{\em Processes underlying long-lived memory and reverberatory
dynamics.}
When the stimulus is withdrawn from the MAPK cascade, the decline in
MAP Kinase activity comes about through
MAPK$^{**}$ binding to MAPK P'ase which dephosphorylates it, resulting
in an increased concentration of MAPK$^*$ [Step $1$,
Figs.~\ref{fig:fig1}(a) and \ref{fig:fig4}(a)]. 
In turn, the phosphatase binds to MAPK$^*$ thereby deactivating it to
MAPK which results in an extremely rapid decline in the concentration of
MAPK$^*$ (Step $2$). Concurrently, the deactivation of MAP2K$^{**}$ is
delayed, as most
of it is bound in the complex MAP2K$^{**}$.MAPK that has a long 
time-scale of disassociation.
To proceed further we can analyze the constituent processes in terms
of the normalized chemical flux $N_{\rm Flux}$ of a molecular species,
i.e., its rate of growth expressed relative to the maximum 
rate of growth of MAPK$^{**}$.
We observe that the suppression of MAP2K$^{**}$ deactivation mentioned
above
results in its normalized chemical flux
exceeding that of MAPK [Fig.~\ref{fig:fig4}(b)].
Thus, there is a net growth in activity in the MAP Kinase layer as 
whenever MAP2K$^{**}$ is released from the complex,
it is available to phosphorylate MAPK which results in an increase in
the concentration of MAPK$^{*}$ (Step $3$). The resulting rise in
MAPK$^{*}$ manifests as a spike in its concentration
[Fig.~\ref{fig:fig4}(a)], and it subsequently gets phosphorylated again to
increase MAPK$^{**}$ concentration even in the absence of
any stimulation (Step $4$).
When the net difference between the normalized flux of MAP2K$^{**}$
and MAPK
reaches a maximum, the normalized
chemical flux of MAPK$^{**}$ attains its highest value
and consequently peak activity of MAP Kinase is observed
[Fig.~\ref{fig:fig4}(c)].
Thus, steps 1-4 represent one complete cycle of MAP Kinase
reverberatory activity characterized
by an initial decline and a subsequent rise in MAPK$^{**}$
concentration. These
steps are subsequently repeated a number of times resulting in
a series of spikes in MAPK activity
[Fig.~\ref{fig:fig4}(d)].
The abrupt nature of the rise and fall of MAP Kinase activity that
manifests as spikes is a consequence of
the bistable nature of the dynamics in the 
MAPK layer of the cascade~\cite{Kholodenko2004,KholodenkoBS2006}. 
We note that similar spiking behavior is also observed in the activity
of MAP2K, with the phase of the MAP2K$^{**}$ spikes shifted slightly
forward with respect to the corresponding ones in MAPK$^{**}$, which
suggests that they result from retrograde propagation of activity from
the MAPK to the MAP2K layer~\cite{Ventura2008}. On the other hand, 
MAP3K shows a monotonic decline in its activity following the removal of the
stimulus.

\begin{figure} [hbt!] 
\begin{center}
\includegraphics[]{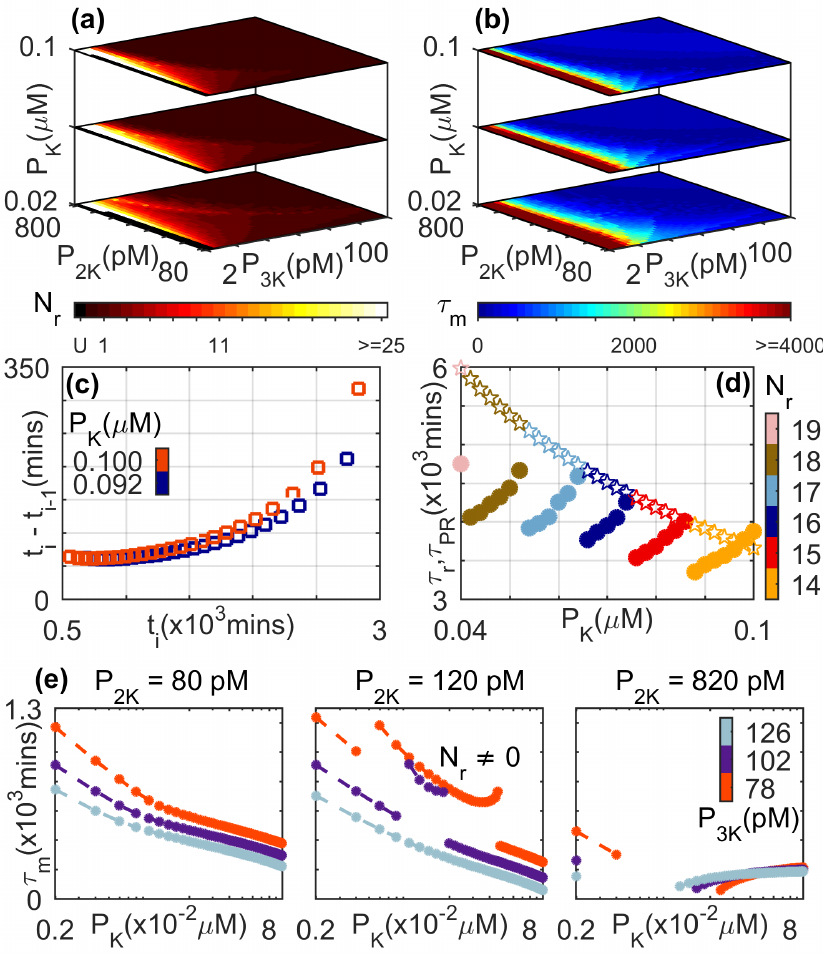}
\end{center}
\caption{Dependence of reverberatory activity on the total
concentrations of the phosphatases MAPK P'ase ($[P_{K}]$), MAP2K P'ase,
($[P_{2K}]$) and MAP3K P'ase ($[P_{3K}]$). (a) The number of spikes $N_{r}$ and (b) the total memory time $\tau_m$ (in minutes) observed on withdrawing an applied stimulus
of amplitude $S$ $=0.8 \times 10^{-6} \mu M$. Situations where the
primary recovery time is longer than a maximum or cut-off value (see
Methods), such that the duration of the reverberatory dynamics cannot
be properly measured, are indicated by the color corresponding to ``U''.
(c) The interval between successive spikes $i-1$ and $i$ increases
with time ($t_{i}$ being the time of occurrence of the $i$th spike).
As the MAPK P'ase concentration is increased, the durations of these
intervals are seen to increase. The total concentrations of the other
two phosphatases are maintained at $[P_{2K}] = 680 pM$ and $[P_{3K}] = 10 pM$.
(d) The variation of primary recovery time $\tau_{\rm PR}$
(stars) and the total duration of reverberatory activity $\tau_{r}$
(filled circles) as a function of total MAPK P'ase concentration are shown
for different values of $N_r$ (indicated by the color bar). 
While $\tau_{\rm PR}$ decreases monotonically with increasing
$[P_{K}]$, $\tau_{r}$ shows a more complex dependence ($[P_{2K}] = 200
pM$ and $[P_{3K}] = 6 pM$). 
(e) Dependence of the total memory time $\tau_{m}$ on total MAPK P'ase
concentration ($[P_K]$ shown in log scale) for
different total concentrations of MAP2K P'ase (values indicated above
each of the three panels) and MAP3K P'ase
(indicated using different colors as shown in the color bar). 
Note that we consider only situations where the system attains a
steady state on
maintaining stimulation. For details of all other parameter values see
Supplementary Information.
}
\label{fig:fig7}
\end{figure}

In order to characterize in detail the memory of prior activity
retained by the cascade which is manifested as
long-lived transient reverberations following the withdrawal of stimulus,
we use the following measures (see Methods): (i)
the primary recovery time ($\tau_{\rm PR}$), (ii) the number of
spikes ($N_{r}$) that occur during the relaxation process, (iii) the temporal
intervals between successive spikes ($t_{i}-t_{i-1}$, where $t_i$ is
the time of occurrence of the $i$th spike event) and (iv) the
total duration of reverberatory activity ($\tau_{r}$) following
primary recovery. The total memory time ($\tau_{m}$) is the sum of
$\tau_{\rm PR}$ and $\tau_{r}$ as indicated in
Fig.~\ref{fig:fig1}~(b).
In the following we use these measures to present a detailed 
characterization of the behavior of the cascade components
over a range of parameter values 
(Figs.~\ref{fig:fig5}-\ref{fig:fig7}).

{\em MAP Kinase cascade components have different recovery
timescales.}
As mentioned earlier, the emergence of long-lived reverberatory
activity of MAPK following the withdrawal of an applied stimulus 
can be linked to the flux imbalance of different
cascade components which suggests 
significant differences in their rates of relaxation.
As shown in Fig.~\ref{fig:fig5}~(a), this is indeed the case, even for
parameter regimes where no spiking activity of
MAPK is observed (i.e., $N_{r} = 0$). 
As can be seen, the nature of increase of the relaxation time with
increasing total concentrations of kinase protein MAP2K is distinct
for the different molecular species and also depends on the state of
their phosphorylation. In the lower layers of the cascade, we also find a
crossover between two regimes seen at lower and higher values of
$[2K]_{tot}$ respectively. These regimes are characterized by
relatively slow and rapid increases (respectively) in the recovery times with
increasing $[2K]_{tot}$, and appear to be related to the steady-state
value attained by MAPK activity upon sustained stimulation of the
cascade for the corresponding value of $[2K]_{tot}$
[Fig.~\ref{fig:fig5}~(b)].
The crossover between the two regimes is seen to occur for a value of
$[2K]_{tot}$ for which $\sim 17\%$ of MAPK is activated for the
parameter values used in Fig.~\ref{fig:fig5}~(b).

The distinct regimes are also observed in the
dependence of the primary recovery time $\tau_{\rm PR}$ on
$[2K]_{tot}$ [Fig.~\ref{fig:fig5}(c)]. As can be observed, 
the difference between the regimes
becomes more pronounced with an increase in the total concentration of
MAP3K. An important point to note is that for lower values of
$[2K]_{tot}$, the recovery time decreases with increasing $[3K]_{tot}$
while the reverse trend is seen for higher values of $[2K]_{tot}$. 
We have verified that increasing the stimulus amplitude $S$
while keeping the total MAP3K concentration fixed has a similar effect
on the relaxation behavior of activated MAPK (see Supplementary
Information). As increasing total concentration of MAP2K results in
increased steady-state activity of MAPK, we conclude that, in
general, higher activity states of MAPK are associated with increasing
relaxation time when either the signal or the substrate (MAP3K) is
increased. Conversely, for states characterized by much lower MAPK activity, 
larger values of $S$ or
$[3K]_{tot}$ results in reduced relaxation periods.

{\em Dependence of reverberatory activity on total kinase
concentrations.}
Diverse cellular environments are characterized by different
total concentrations of the various molecular components of the MAPK
cascade. Thus, in order to determine
the robustness of spiking and reverberatory
activity following the removal of an applied stimulus, 
it is important to see how they are 
affected by varying total kinase concentrations. Such a study will also
indicate the ease with which these phenomena can be experimentally
observed. Fig.~\ref{fig:fig6} shows the variation of different
measures of reverberatory activity on the total concentrations of MAPK,
MAP2K and MAP3K. While there is a complex dependence on these
parameters for the exact number of spikes $N_r$ and the
duration of the total memory time $\tau_m$, the phenomenon of
reverberatory activity following withdrawal of stimulation can be
observed
over a large range of the parameter space, underlining its robustness.
We also observe that on increasing $[3K]_{tot}$, the response of
$N_{r}$ to variation in
$[K]_{tot}$ and $[2K]_{tot}$ becomes relatively homogeneous.
Increasing the stimulus amplitude $S$ [compare panels (a,c,e)
with (b,d,f) of Fig.~\ref{fig:fig6}] does not seem to alter the
qualitative nature of the variation in $N_r$ and $\tau_m$ over the
parameter space in general, although we do observe that the domains
corresponding to different values of $N_r$ occupy different regions
[Fig.~\ref{fig:fig6}(e and f)]. 
Note that for low $[3K]_{tot}$, high values of $N_{r}$ are observed
to coexist with low values of $\tau_{m}$ [Fig.~\ref{fig:fig6}(a,c and
b,d)]. While it may appear surprising that these two measures of
memory are not in consonance in this region of parameter space, 
it can be explained by
noting that the stimulated system is in an
oscillatory state, and following the removal of the signal
these relatively high-frequency oscillations cease after a short
duration. 
Fig.~\ref{fig:fig6}~(g) suggests that the variation seen in $\tau_m$
as a function of the total MAPK concentration for a specific $N_r$ is
mostly governed by $\tau_r$, the total duration of reverberatory
activity, with the corresponding dependence of $\tau_{PR}$ on $[K]_{tot}$
being weak.

As the total MAPK concentration is increased, we observe that while
the primary recovery time increases almost linearly, the nature of the
reverberatory dynamics as reflected in $\tau_r$ shows a more complex
dependence on $[K]_{tot}$ [Fig.~\ref{fig:fig6}~(g)]. If for a given
value of $[K]_{tot}$ the MAPK activity following withdrawal of the
stimulus shows $N_r$ spikes over a duration of $\tau_r$, then on increasing $[K]_{tot}$ the time-interval between the spikes increases (thereby resulting in an increase of $\tau_r$) until a critical value beyond which the last of the $N_r$ spike no longer appears. Thus, at this point $N_r$ reduces by unity with a concomitant drop in $\tau_r$. This series of events is repeated for steadily decreasing values of $N_r$ as the total MAPK concentration is increased further. Each value of $N_r$ is associated with a characteristic rate of increase in $\tau_r$ with $[K]_{tot}$. With a reduction in $N_r$ (as a result of increasing $[K]_{tot}$), this rate is found to decrease as well, which suggests a saturation of the system response. These results are robust with respect to different choices of total MAP3K concentration as can be seen from
Fig.~\ref{fig:fig6}(h), suggesting that similar behavior will be seen for a range of strengths for the applied signal (see Supplementary Information). 

{\em Dependence of reverberatory activity on total phosphatase
concentrations.}
We have also investigated the role that phosphatase
availability plays on the reverberatory activity of the cascade following the withdrawal of the stimulus.
As is the case for total kinase concentrations shown in
Fig.~\ref{fig:fig6}, we see from Fig.~\ref{fig:fig7}~(a-b) that the
number of spikes $N_r$ and the duration of total memory
time $\tau_m$ depend on the total concentrations of the phosphatases
MAPK P'ase, MAP2K P'ase,
and MAP3K P'ase.
For larger values of the concentrations, viz., $[P_K]$,
$[P_{2K}]$ and $[P_{3K}]$, respectively, 
the system operates in the low-amplitude response
regime. As mentioned earlier, the reverberatory
MAPK dynamics during recovery following withdrawal of the
applied stimulus will not be seen in this regime. 
As the phosphatase concentrations are decreased, spiking behavior of
MAPK activity
is observed with both $\tau_m$ and $N_r$ attaining high values in an
optimal range.
The large variation seen in $\tau_m$ [Fig.~\ref{fig:fig7}~(b)] arises
as regions in $[P_{2K}]$-$[P_{3K}]$ parameter
space characterized by the same value of $N_{r}$ are seen to
exhibit a range of different values of $\tau_{r}$ and $\tau_{PR}$
[Fig.~\ref{fig:fig7}~(d)].
For reverberatory activity associated with a specific $N_{r}$, we
observe that the duration $\tau_{r}$ increases with increasing total
MAPK P'ase concentration. This is a consequence of
the intervals between successive spikes ($t_{i}-t_{i-1}$) increasing
with $[P_{K}]$ as
is shown in Fig.~\ref{fig:fig7}~(c). Note that the results are
qualitatively similar for different amplitudes of the applied
stimulus (see
Supplementary Information).
However, increasing $[P_K]$ results also in decreased time for
primary recovery $\tau_{PR}$ [Fig.~\ref{fig:fig7}~(d)], which in
conjunction with the previously mentioned result leads to
non-monotonic dependence of the total memory time $\tau_{m}$ on
phosphatase availability. While this
non-monotonicity is suggested in Fig.~\ref{fig:fig7}~(b), it is shown clearly in
Fig.~\ref{fig:fig7}~(e) where the central panel corresponds
to situations where spiking behavior is observed in MAPK
activity.
Investigation into the
dependence of $\tau_{m}$ on $P_{K}$
[Fig.~\ref{fig:fig7}(e)] reveals that the range of $[P_K]$ over which
reverberatory activity (i.e., $N_r \neq 0$) occurs is demarcated by
discontinuities in the functional dependence of 
$\tau_{m}$ on $P_{K}$. For intermediate $P_{2K}$
[Fig.~\ref{fig:fig7}(e), central panel] where the system attains a
steady state on maintaining stimulation, the spiking
activity following withdrawal of the stimulus becomes more prominent for low total concentration of MAP3K
P'ase. For higher $P_{2K}$
[Fig.~\ref{fig:fig7}(e), right panel] where the system becomes
oscillatory over an intermediate range of $[P_K]$, 
reverberatory activity is observed over a broader range of $[P_{3K}]$.
While we have assumed that the same phosphatase
acts on both the singly and doubly phosphorylated forms of the
kinase in a particular layer of the cascade (as in the canonical
Huang-Ferrell model), we have explicitly verified that our results are
not sensitively dependent on this. 

\section{Discussion} 
In this paper we have shown that an isolated MAPK signaling module
can serve as a fundamental motif in the intra-cellular signaling network
for imparting a form of short-term memory to the cell.
The emergence of long-lived reverberatory activity reported
here arises from the diversity of relaxation timescales for the
different components of the MAP Kinase cascade, which results in flux imbalance
between activation of the MAPK layer and deactivation in the MAP2K
layer. One may therefore expect to observe results qualitatively
similar to what has been reported here
whenever the system has disparate timescales
regardless of the actual molecular concentrations and kinetic rates
which can vary substantially across different
cells~\cite{Stumpf2016,Legewie2013,Das2013}.
Thus, as the MAPK cascade is present in all eukaryotic
cells~\cite{Seger1995,Johnson1999}, the
mechanism for short-term memory in such a signaling cascade that
is presented here may hold for such cells in
general.
As the duration of MAPK$^{**}$ activity is critical for many cellular
decisions~\cite{Marshall1995}, e.g., the prolonged activation of ERK
resulting in its translocation to the nucleus~\cite{Seger2011}, the
persistent reverberatory activity seen here may play a non-trivial
role in regulation of cellular functions.

The basal level activity of MAPK in a normal cell is maintained at a
low proportion of the total MAPK 
concentration and serves several biological functions~\cite{Kosik1998}. 
We observe a crossover between two qualitatively distinct regimes of
relaxation behavior of MAPK$^{**}$ occurring at a steady state that is
characterized by relatively low proportion of activation of the
available MAPK [$\sim 17\%$ in Fig.~\ref{fig:fig5}~(b)]. Thus, there
appears to be
an effective threshold for MAPK activity (which may be related
to its basal state level) that demarcates the different 
relaxation regimes following the removal of the applied stimulus. 
A similar crossover is also observed for the 
primary recovery time $\tau_{PR}$.

It is known that ERK MAPK isoforms (e.g., p42 and p44) are
abundantly expressed in non-dividing terminally
differentiated neurons~\cite{Sweatt2001}. 
Activation of MAPK by spaced stimulation is
known to be responsible for
morphological changes in dendrites~\cite{Tsien2001}. Studies also
suggest that the activation of the MAPK pathway is linked with
associative learning in the mammalian nervous
system, synaptic plasticity and
neurological
memory~\cite{Carew2013,Tsien2001,Carew2008,Sweatt1998,Sweatt2001}.
An intriguing possibility suggested by the results reported here is
that the observed repeated spiking in MAPK activity
may function as an effective temporally spaced signal to the nucleus
of a neuron. This can then 
facilitate subsequent
changes in the cell required for memory formation.

Another well-known example of eukaryotic cellular memory is observed
during chemotactic migration along the gradient of a chemical 
signal~\cite{Rappel2014,Shah2016}.
The directionality of migration is known to persist for a
certain duration, even if the chemical gradient is altered or becomes
static. Studies show that the protein Moesin contributes to the long-lived
rigidity of the cytoskeleton assembly that subsequently leads to the
directional memory in polarized migrating cells~\cite{Shah2016}.
However, the intra-cellular processes that underlie the persistent
activity of Moesin in the absence of a gradient
mediated signal are still largely unknown. Evidence suggests that the
regulation of Moesin and other ERM
proteins are linked with the activity of the MAPK
pathway~\cite{Huang2009,Wang2006}. The long-term
reverberatory activity of MAPK following the withdrawal of a stimulus
that is reported here may be a possible mechanism underlying such
persistent cellular behavior.
To conclude, we have shown the possibility of long-lived reverberatory activity
in a signaling cascade following the withdrawal of external stimuli.
Our results suggest
a mechanism through which the intra-cellular signaling system can
encode short-term memory of signals to which the cell was previously exposed.
The large-amplitude spiking activity of MAPK following the removal of
a prior stimulus may also provide a mechanism for signal integration and
learning when the cascade is repeatedly stimulated.
We note that there may be additional factors not considered
here that may lengthen the
persistence of reverberatory activity, including scaffold proteins
that increase the lifetime of kinase complexes. Our
results suggest that the MAPK cascade potentially has a key role in
shaping the information processing capabilities
of eukaryotic cells in diverse environments.
   
%
%
%
%

\begin{acknowledgements}
{\small SNM is supported by the IMSc Complex Systems Project ($12^{\rm
th}$ Plan). The simulations required for this work
were done in the High Performance Computing facility (Nandadevi and
Satpura) of IMSc which is partially
funded by DST (SR/NM/NS-44/2009). 
We thank James Ferrell, Upinder 
Bhalla, Tharmaraj Jesan, Uddipan
Sarma, Bhaskar Saha, Jose Faro, Vineeta Bal, J.
Krishnan, Mukund Thattai,
Marsha Rosner and Pamela Silver for helpful discussions.}
\end{acknowledgements}

\pagebreak

\onecolumngrid

\newpage
\begin{center}
{\large \bf SUPPLEMENTARY INFORMATION}
\end{center}

        \setcounter{section}{0}
        \renewcommand{\thesection}{S\arabic{section}}%
        \setcounter{table}{0}
        \renewcommand{\thetable}{S\arabic{table}}%
        \setcounter{figure}{0}
        \renewcommand{\thefigure}{S\arabic{figure}}%
        \setcounter{equation}{0}
        \renewcommand{\theequation}{S\arabic{equation}}%



\section{The Model Equations}

\noindent 


\begin{table}[ht]
\caption{Components of the MAPK Cascade} 
\vskip 0.18cm
\centering 
\begin{tabular}{l l l} 
\hline\hline 
Component & Notation & Symbol\\ [0.5ex] 
\hline 
Mitogen-activated Protein Kinase Kinase Kinase & MAP3K & 3K \\ 
Singly Phosphorylated Mitogen-activated Protein Kinase Kinase Kinase & MAP3K* & 3K* \\
Mitogen-activated Protein Kinase Kinase & MAP2K & 2K \\
Singly Phosphorylated Mitogen-activated Protein Kinase Kinase & MAP2K* & 2K* \\
Doubly Phosphorylated Mitogen-activated Protein Kinase Kinase & MAP2K** & 2K** \\
Mitogen-activated Protein Kinase & MAPK & K \\
Singly Phosphorylated Mitogen-activated Protein Kinase & MAPK* & K* \\
Doubly Phosphorylated Mitogen-activated Protein Kinase & MAPK** & K** \\ 
MAP3K-Phosphatase & 3K P'ase & P$_{\rm 3K}$ \\ 
MAP2K-Phosphatase & 2K P'ase & P$_{\rm 2K}$ \\
MAPK-Phosphatase & K P'ase & P$_{\rm K}$ \\ [1ex]
\hline 
\end{tabular}
\end{table}

\vskip 0.5cm

\noindent
\textbf{The three layer MAPK cascade comprises the following
enzyme-substrate reactions:}

\begin{equation*} 
S + 3K\begin{array}{c} {\stackrel{k_{1} }{\longrightarrow} } \\ {\xleftarrow[{k_{-1} }]{} } \end{array}S.3K\stackrel{k_{2} }{\longrightarrow} S + 3K^{*} 
\end{equation*} 

\begin{equation*} 
P_{3K} + 3K^{*}\begin{array}{c} {\stackrel{kp_{1} }{\longrightarrow} } \\ {\xleftarrow[{kp_{-1} }]{} } \end{array}3K^{*}.P_{3K}\stackrel{kp_{2} }{\longrightarrow} P_{3K} + 3K 
\end{equation*} 

\begin{equation*} 
3K^{*} + 2K\begin{array}{c} {\stackrel{k_{3} }{\longrightarrow} } \\ {\xleftarrow[{k_{-3} }]{} } \end{array}3K^{*}.2K\stackrel{k_{4} }{\longrightarrow} 3K^{*} + 2K^{*} 
\end{equation*} 

\begin{equation*} 
P_{2K} + 2K^{*}\begin{array}{c} {\stackrel{kp_{3} }{\longrightarrow} } \\ {\xleftarrow[{kp_{-3} }]{} } \end{array}2K^{*}.P_{2K}\stackrel{kp_{4} }{\longrightarrow} P_{2K} + 2K 
\end{equation*} 

\begin{equation*} 
3K^{*} + 2K^{*}\begin{array}{c} {\stackrel{k_{5} }{\longrightarrow} } \\ {\xleftarrow[{k_{-5} }]{} } \end{array}3K^{*}.2K^{*}\stackrel{k_{6} }{\longrightarrow} 3K^{*} + 2K^{**} 
\end{equation*} 

\begin{equation*} 
P_{2K} + 2K^{**}\begin{array}{c} {\stackrel{kp_{5} }{\longrightarrow} } \\ {\xleftarrow[{kp_{-5} }]{} } \end{array}2K^{**}.P_{2K}\stackrel{kp_{6} }{\longrightarrow} P_{2K} + 2K^{*} 
\end{equation*} 

\begin{equation*} 
2K^{**} + K\begin{array}{c} {\stackrel{k_{7} }{\longrightarrow} } \\ {\xleftarrow[{k_{-7} }]{} } \end{array}2K^{**}.K\stackrel{k_{8} }{\longrightarrow} 2K^{**} + K^{*} 
\end{equation*} 

\begin{equation*} 
P_{K} + K^{*}\begin{array}{c} {\stackrel{kp_{7} }{\longrightarrow} } \\ {\xleftarrow[{kp_{-7} }]{} } \end{array}K^{*}.P_{K}\stackrel{kp_{8} }{\longrightarrow} P_{K} + K 
\end{equation*} 

\begin{equation*} 
2K^{**} + K^{*}\begin{array}{c} {\stackrel{k_{9} }{\longrightarrow} } \\ {\xleftarrow[{k_{-9} }]{} } \end{array}2K^{**}.K^{*}\stackrel{k_{10} }{\longrightarrow} 2K^{**} + K^{**} 
\end{equation*} 

\begin{equation*} 
P_{K} + K^{**}\begin{array}{c} {\stackrel{kp_{9} }{\longrightarrow} } \\ {\xleftarrow[{kp_{-9} }]{} } \end{array}K^{**}.P_{K}\stackrel{kp_{10} }{\longrightarrow} P_{K} + K^{*}
\end{equation*} 


\noindent
\textbf{The above enzyme-substrate reactions can be expressed in terms
of the following coupled ordinary differential equations (ODEs):}


\begin{eqnarray*}
\dt{\FF1}  &=&    \LL1.\FF2 + \LL2.\FF3 - \LL3.\br{S}.\FF1 \,,\\
\dt{\FF2}  &=&    \LL3.\br{S}.\FF1 - (\LL1 + \LL4).\FF2 \,,\\
\dt{\FF3}  &=&    \LL5.\GG1.\FF4  - (\LL2 + \LL6).\FF3\,,\\
\dt{\FF4}  &=&    \LL4.\FF2 + \LL6.\FF3 - \LL5.\GG1.\FF4\\
           & & + (\LL7  + \LL8).\FF6   - \LL9. \FF4.\FF5\\
           & & + (\LL10 + \LL11).\FF9  - \LL12.\FF4.\FF8\,,\\
\dt{\FF5}  &=&    \LL7.\FF6 + \LL19.\FF7 - \LL9.\FF4.\FF5\,,\\
\dt{\FF6}  &=&    \LL9.\FF4.\FF5  - (\LL7 + \LL8).\FF6\,,\\
\dt{\FF7}  &=&    \LL20.\GG2.\FF8 - (\LL19 + \LL21).\FF7\,,\\
\dt{\FF8}  &=&    \LL8.\FF6 + \LL21.\FF7 - \LL20.\GG2.\FF8 \\
           & &  + \LL10.\FF9 - \LL12.\FF4.\FF8 + \LL23.\FF10\,,\\
\dt{\FF9}  &=&    \LL12.\FF4.\FF8  - (\LL11 + \LL10).\FF9\,,\\
\dt{\FF10} &=&    \LL22.\GG3.\FF11 - (\LL23 + \LL24).\FF10\,,\\
\dt{\FF11} &=&    \LL11.\FF9 + \LL24.\FF10 - \LL22.\GG3.\FF11\\
           & & + (\LL25 + \LL26).\FF13 - \LL27.\FF11.\FF12\\
           & & + (\LL28 + \LL29).\FF16 - \LL30.\FF11.\FF15\,,\\
\dt{\FF12} &=&    \LL25.\FF13 + \LL31.\FF14 - \LL27.\FF11.\FF12\,,\\
\dt{\FF13} &=&    \LL27.\FF11.\FF12 - (\LL26 + \LL25).\FF13\,,\\
\dt{\FF14} &=&    \LL32.\GG4.\FF15  - (\LL33 + \LL31).\FF14\,,\\
\dt{\FF15} &=&    \LL26.\FF13 + \LL33.\FF14 -\LL32.\GG4.\FF15\\
           & &  + \LL28.\FF16 - \LL30.\FF11.\FF15 + \LL34.\FF17\,,\\
\dt{\FF16} &=&    \LL30.\FF11.\FF15 - (\LL28 + \LL29).\FF16\,,\\
\dt{\FF17} &=&    \LL35.\GG5.\FF18 - (\LL36 + \LL34).\FF17\,,\\
\dt{\FF18} &=&   \LL29.\FF16 + \LL36.\FF17 - \LL35.\GG5.\FF18\,.
\end{eqnarray*}

\noindent where
\begin{align*}
\br{S} &= \br{S}_{\rm tot}  - \FF2\,,\;\;\\
\GG1   &= [P_{3K}] - \FF3\,,\;\; \\
\GG2   &= [P_{2K}] - \FF7\ - \FF10\,,\;\; \\
\GG4   &= [P_{K}] - \FF14\ - \FF17\,.
\end{align*}

\noindent It is explicitly ensured that the total concentrations of
all individual kinases and phosphatases are conserved at all times. 
The concentrations of the different molecular species can vary over
several orders of magnitudes. We have therefore numerically solved the
equations using low relative and absolute tolerances in order to
ensure the accuracy of the resulting time-series.


\section{System Parameters}

The numerical values for the reaction rates are obtained from
Ref.~[13], and are listed in Table~\ref{table:Reaction Rates}.


\begin{table}[ht]
\caption{Reaction Rates}
\vskip 0.15cm
\centering 
\begin{tabular}{l l l} 
\hline\hline 
Rate constant & Value & Units \\ [0.5ex] 
\hline 
$k_{1}$ & 1002 & $(\mu M.{\rm min})^{-1}$ \\ 
$k_{-1}$ & 150 & ${\rm min}^{-1}$ \\
$k_{2}$ & 150 & ${\rm min}^{-1}$  \\
$kp_{1}$ & 1002 & $(\mu M.{\rm min})^{-1}$ \\
$kp_{-1}$ & 150 & ${\rm min}^{-1}$ \\ 
$kp_{2}$ & 150 & ${\rm min}^{-1}$ \\
$k_{3}$ & 1002 & $(\mu M.{\rm min})^{-1}$  \\
$k_{-3}$ & 30 & ${\rm min}^{-1}$ \\
$k_{4}$ & 30 & ${\rm min}^{-1}$ \\ 
$kp_{3}$ & 1002 & $(\mu M.{\rm min})^{-1}$ \\
$kp_{-3}$ & 150 & ${\rm min}^{-1}$  \\
$kp_{4}$ & 150 & ${\rm min}^{-1}$ \\
$k_{5}$ & 1002 & $(\mu M.{\rm min})^{-1}$ \\ 
$k_{-5}$ & 30 & ${\rm min}^{-1}$ \\
$k_{6}$ & 30 & ${\rm min}^{-1}$  \\
$kp_{5}$ & 1002 & $(\mu M.{\rm min})^{-1}$ \\
$kp_{-5}$ & 150 & ${\rm min}^{-1}$ \\ 
$kp_{6}$ & 150 & ${\rm min}^{-1}$ \\
$k_{7}$ & 1002 & $(\mu M.{\rm min})^{-1}$  \\
$k_{-7}$ & 30 & ${\rm min}^{-1}$ \\
$k_{8}$ & 30 & ${\rm min}^{-1}$ \\ 
$kp_{7}$ & 1002 & $(\mu M.{\rm min})^{-1}$ \\ 
$kp_{-7}$ & 150 & ${\rm min}^{-1}$ \\
$kp_{8}$ & 150 & ${\rm min}^{-1}$  \\
$k_{9}$ & 1002 & $(\mu M.{\rm min})^{-1}$ \\
$k_{-9}$ & 150 & ${\rm min}^{-1}$ \\ 
$k_{10}$ & 150 & ${\rm min}^{-1}$ \\
$kp_{9}$ & 1002 & $(\mu M.{\rm min})^{-1}$  \\
$kp_{-9}$ & 150 & ${\rm min}^{-1}$  \\
$kp_{10}$ & 150 & ${\rm min}^{-1}$ \\ [1ex] 
\hline 
\end{tabular}
\label{table:Reaction Rates} 
\end{table}

\newpage

\begin{table}[ht]
\caption{Total concentration (in $\mu M$) of the kinase proteins for
Fig.~4 (panels a--c) and Fig.~7} 
\vskip 0.18cm
\centering 
\begin{tabular}{c c c} 
\hline\hline 
$[K]_{tot}$ & $[2K]_{tot}$ & $[3K]_{tot}$ \\ [0.5ex] 
\hline 
4.8 & 1.2 & 0.0030 \\ [1ex]
\hline 
\end{tabular}
\label{table:kinases47} 
\end{table}

\begin{table}[ht]
\caption{Total concentration (in $\mu M$) of the phosphatase proteins for Figs.~2--6} 
\vskip 0.18cm
\centering 
\begin{tabular}{l l} 
\hline\hline 
Phosphatase Protein& Value \\ [0.5ex] 
\hline 
MAP3K-Phosphatase & $1 \times 10^{-4}$  \\ 
MAP2K-Phosphatase & $3 \times 10^{-4}$  \\
MAPK-Phosphatase & 0.05   \\ [1ex]
\hline 
\end{tabular}
\label{table:phosphatases} 
\end{table}

\begin{table}[ht]
\caption{Total concentration (in $\mu M$) of the kinase proteins for Figs.~2--3} 
\vskip 0.18cm
\centering 
\begin{tabular}{l l l l} 
\hline\hline 
Panels & $[K]_{tot}$ & $[2K]_{tot}$ & $[3K]_{tot}$  \\ [0.5ex] 
\hline 
(a) and (f) & 3.0 & 3.0 & 0.0080 \\ 
(b) and (g) & 1.0 & 2.4 & 0.0024 \\
(c) and (h) & 1.2 & 6.0 & 0.0028 \\
(d) and (i) & 2.0 & 2.2 & 0.0024 \\
(e) and (j) & 4.8 & 6.0 & 0.0014 \\ [1ex]
\hline 
\end{tabular}
\label{table:kinases23} 
\end{table}

\newpage

\section{Supplementary Figures}

\begin{figure*} [hbt!]
\begin{center}
\includegraphics[]{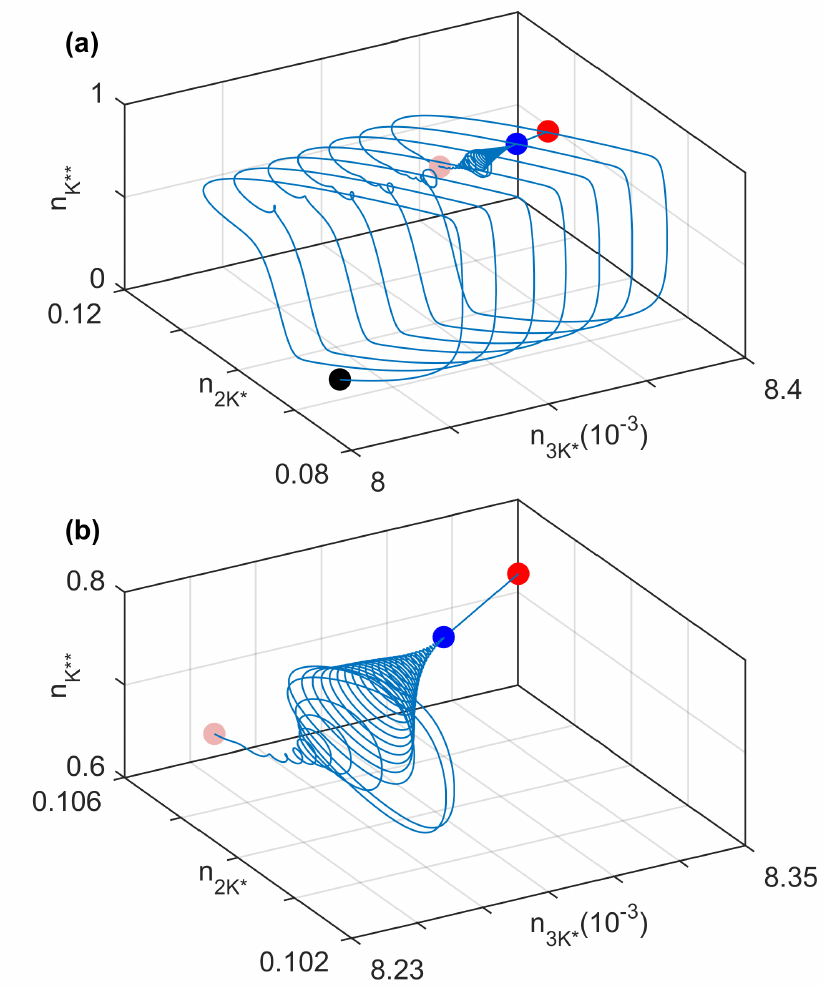}
\end{center}
\caption{Magnified views of the phase space trajectory shown in
Fig.~2(i). The blue markers correspond to the final point in the time
series displayed in Fig.~2(d), while the red markers indicate the
fixed point of the dynamical system in the presence of stimulus. 
(a) Magnified view of the trajectory
beginning from the black marker shown in Fig.~2(i).
The pink marker denotes
the starting point of the segment of the trajectory
displayed in panel (b). (b) Further magnification of the section of phase-plane
trajectory shown in 
(a), corresponding to the duration when the system moves away from the
unstable limit cycle and converges to the stable fixed point.
}
\label{fig:fig8}
\end{figure*}

\begin{figure*}
\begin{center}
\includegraphics[width=0.49\linewidth]{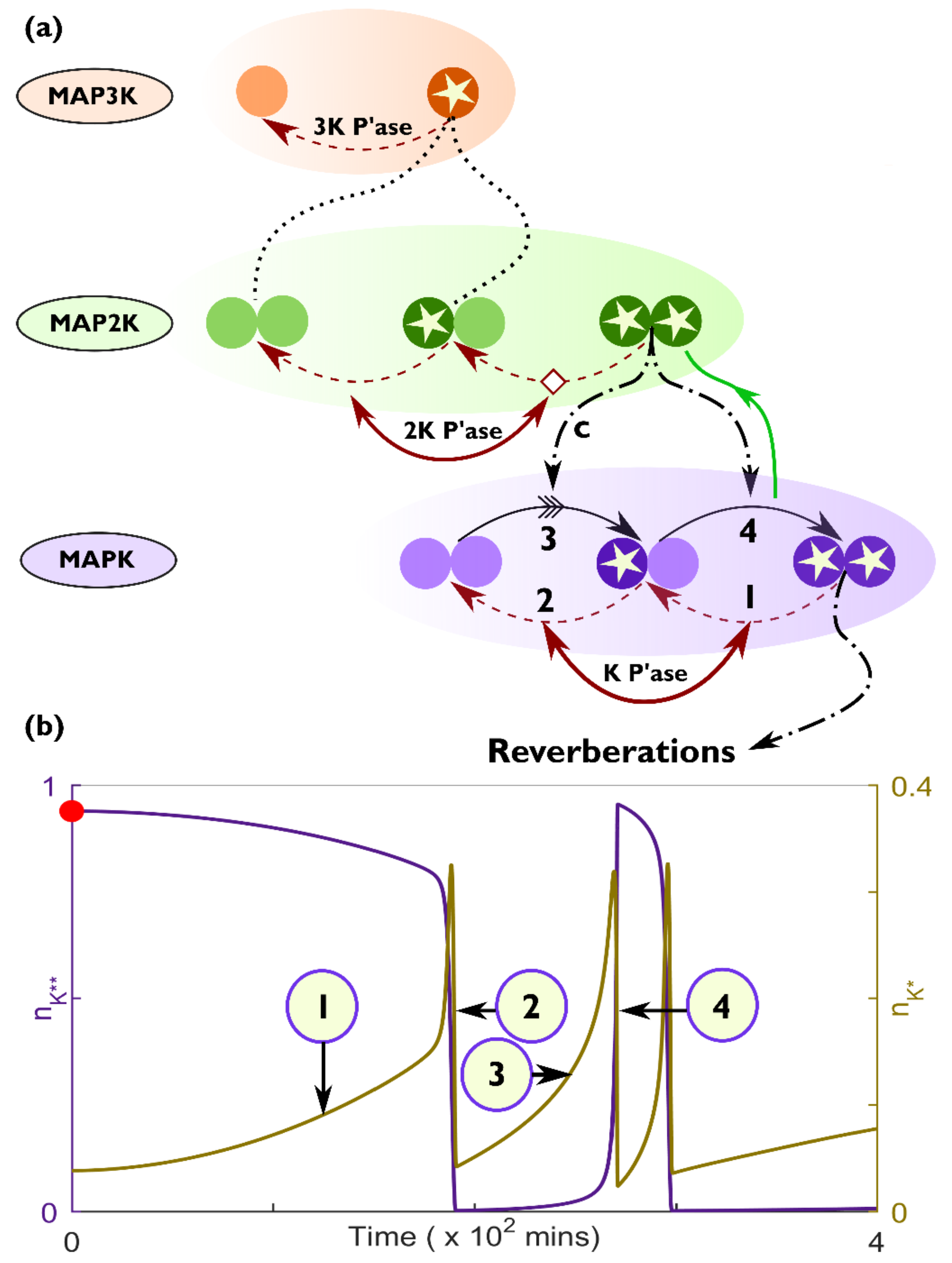}
\end{center}
\caption{Processes underlying long-lived memory and reverberatory
dynamics. (a) Schematic representation of MAPK cascade showing the
processes that occur
subsequent to removing a stimulus.
The numbers ($1-4$) represent the sequence of events that lead to the
emergence of
the post-stimulus large-amplitude spiking activity shown in (b). 
The enzyme-substrate protein complex formed during activation of MAPK by doubly phosphorylated MAP2K is
indicated by ``c''. The green arrow from the MAPK layer to the
MAP2K layer represents the release of doubly phosphorylated MAP2K from
downstream complexes. 
(b) A characteristic
time-series for the normalized concentration of singly and doubly phosphorylated
MAPK ($n_{K^{*}}$ and $n_{K^{**}}$, respectively) following the removal of an applied stimulus of
amplitude $S = 2.0 \times 10^{-6} \mu M$ at $t = 0$. The numbers
($1-4$) represent the same events shown in (a). 
The total concentrations of the kinases and phosphatases used for
generating the time-series are provided in Tables~\ref{table:kinases47} and \ref{table:phosphatases}, respectively.
}
\label{fig:fig9}
\end{figure*}

\begin{figure*} 
\begin{center}
\includegraphics[ ]{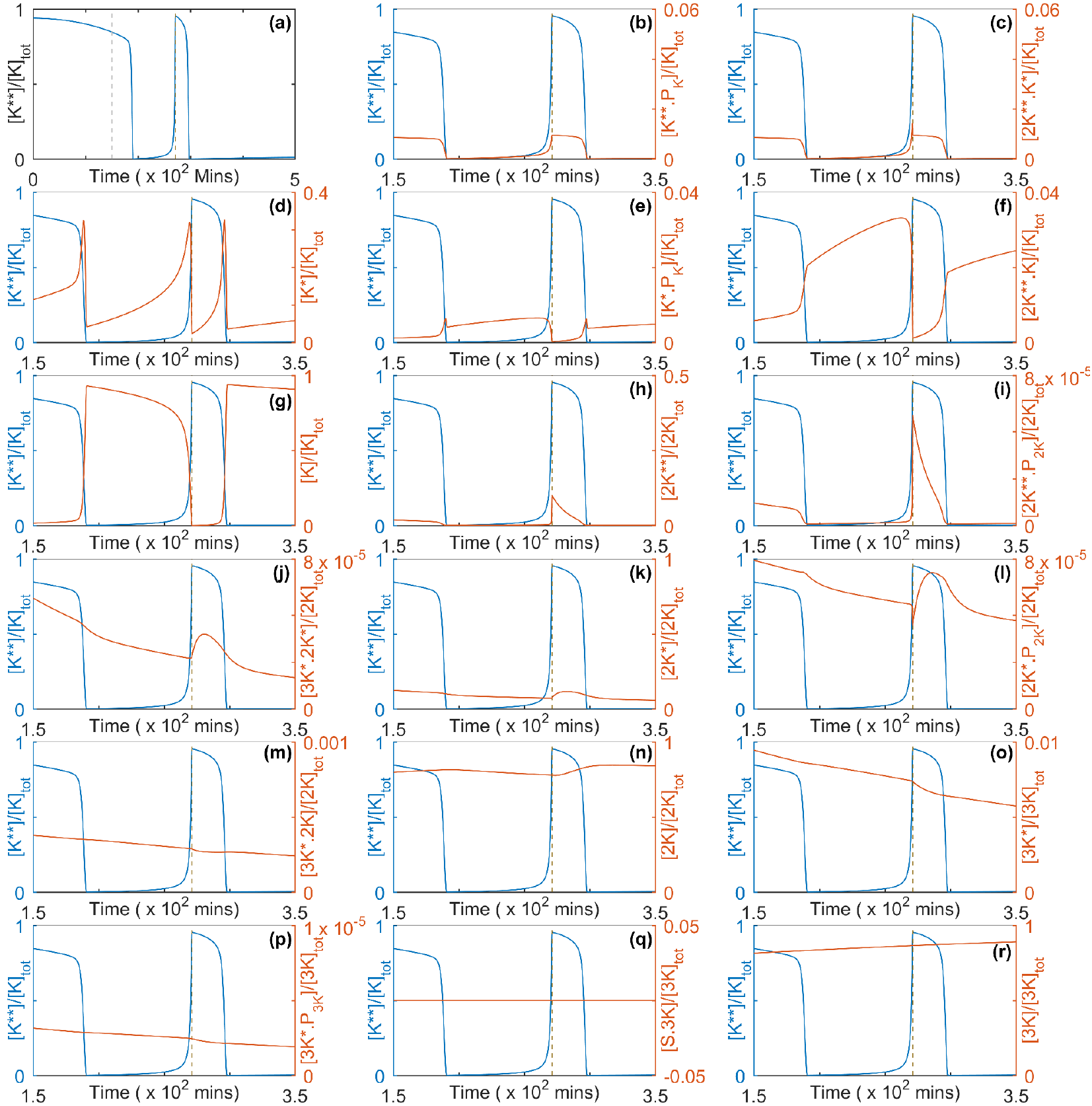}
\end{center}
\caption{Characteristic dynamics of the molecular components of the
MAP Kinase cascade following withdrawal of a
stimulus. (a) The time-series of the normalized concentration of
doubly phosphorylated MAPK ($[K^{**}]/[K]_{tot}$) following removal
of an applied stimulus with amplitude $S = 2.0 \times 10^{-6} \mu M$
at $t = 0$. (b-r) Time-series of the normalized concentrations of the
different components of 
the MAPK cascade, shown starting from $t=150$ minutes after
withdrawing the stimulus, displayed together with the time-series of 
normalized MAPK activity $[K^{**}]/[K]_{tot}$. 
The total concentrations of the kinases
and phosphatases used for generating the figures are provided in
Tables~\ref{table:kinases47} and \ref{table:phosphatases},
respectively.
}
\label{fig:fig10}
\end{figure*}

\begin{figure*}
\begin{center}
\includegraphics[]{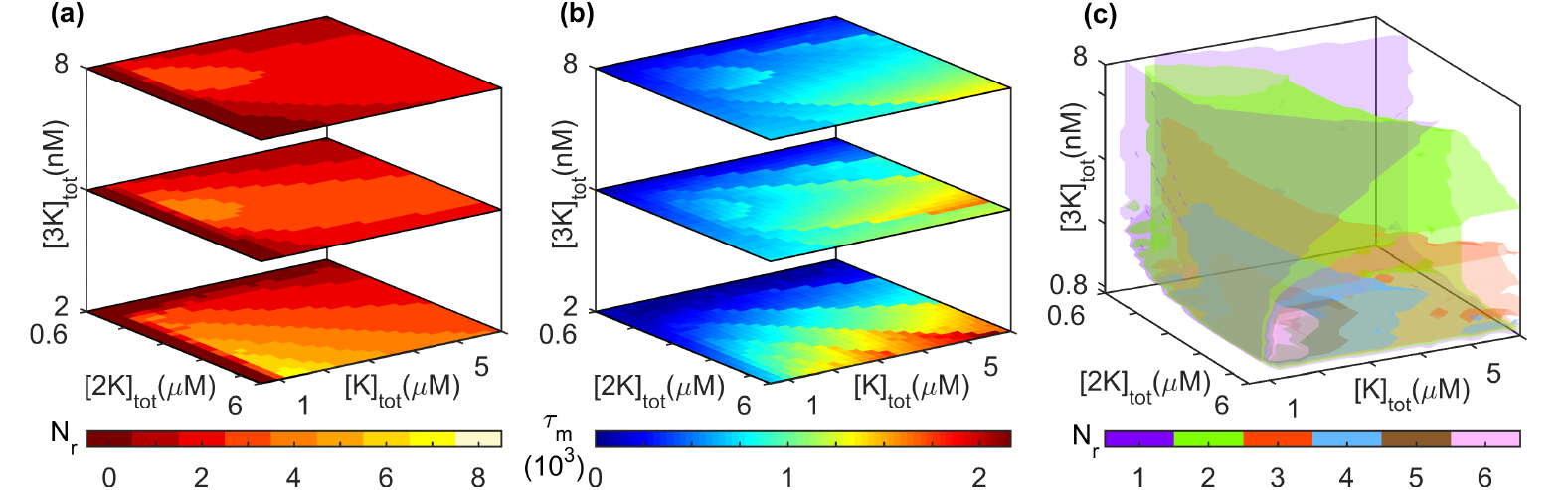}
\end{center}
\caption{Dependence of reverberatory activity on the total kinase
concentrations, viz., MAPK ($[K]_{tot}$),
MAP2K ($[2K]_{tot}$) and MAP3K ($[3K]_{tot}$). 
(a) The number of post-stimulus spikes $N_{r}$, (b) the total
memory time $\tau_{m}$ (in minutes), and (c) isosurfaces for $N_r$ 
observed on withdrawing an applied stimulus
of amplitude $S=2.0 \times 10^{-6} \mu M$, are shown as functions of the three
total kinase concentrations. The total concentrations of the phosphatases are held fixed for (a-c) and are provided in Table~\ref{table:phosphatases}.
}
\label{fig:fig11}
\end{figure*}


\begin{figure*} 
\begin{center}
\includegraphics[ ]{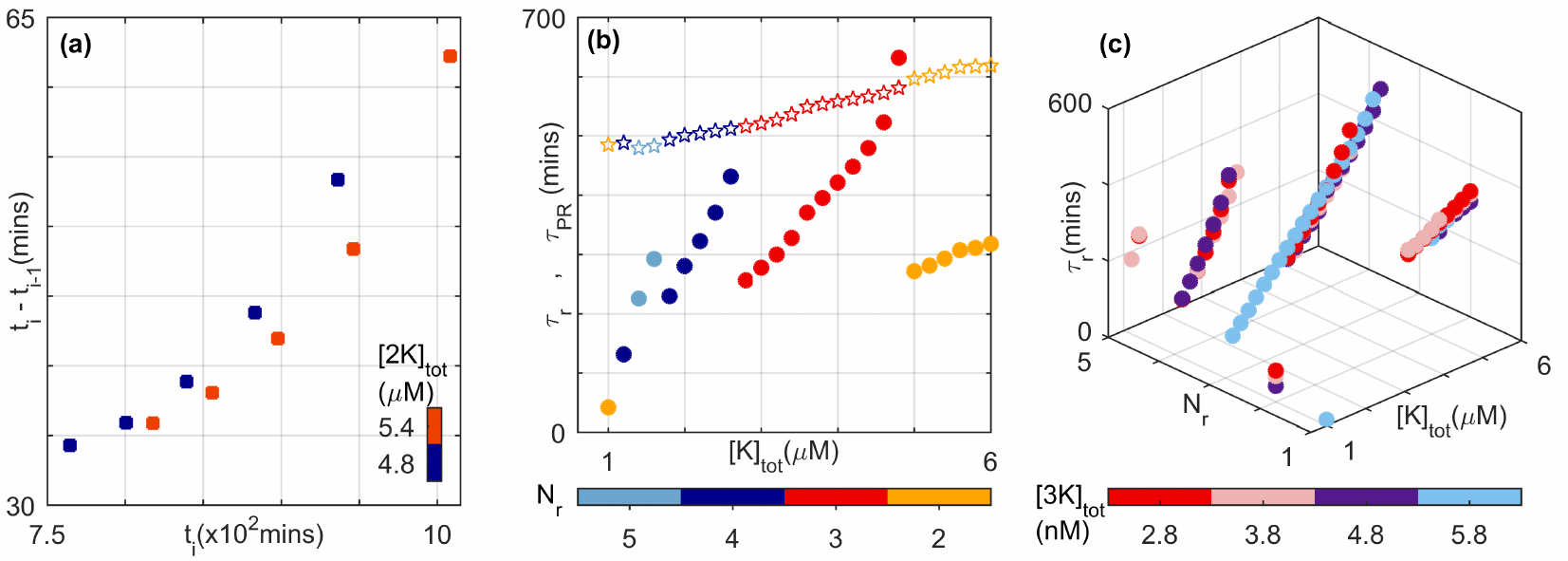}
\end{center}
\caption{Characterization of the reverberatory dynamics observed after
withdrawing a stimulus having
amplitude $S=2.0 \times 10^{-6} \mu M$. 
(a) The interval between successive spikes $i-1$ and $i$ increases
with time ($t_{i}$ being the time of occurrence of the $i$th spike) for two distinct total concentrations of MAP2K.
The total concentrations of MAPK and MAP3K are $[K]_{tot} = 1.2 \mu M$ and $[3K]_{tot} = 2.8 nM$, respectively. 
(b) The primary recovery time $\tau_{\rm PR}$
(stars) and the total duration of reverberatory activity $\tau_{r}$
(filled circles) are shown for different values of $N_r$ (indicated by the color bar). While $\tau_{\rm PR}$ increases monotonically with increasing
total MAPK concentration, $\tau_{r}$ shows a more complex dependence ($[2K]_{tot} = 3 \mu M$ and $[3K]_{tot} = 4
nM$). (c) The dependence of $\tau_{r}$ on $[K]_{tot}$ for different
values of $N_{r}$ has a similar nature for different choices of
$[3K]_{tot}$ (indicated by the color bar, $[2K]_{tot} = 3 \mu M$). Note
that for panel (c), we consider only situations where the
system attains a steady state on maintaining the stimulation.
For the total concentrations of the phosphatases see Table~\ref{table:phosphatases}.
}
\label{fig:fig12}
\end{figure*}

\begin{figure*} 
\begin{center}
\includegraphics[ ]{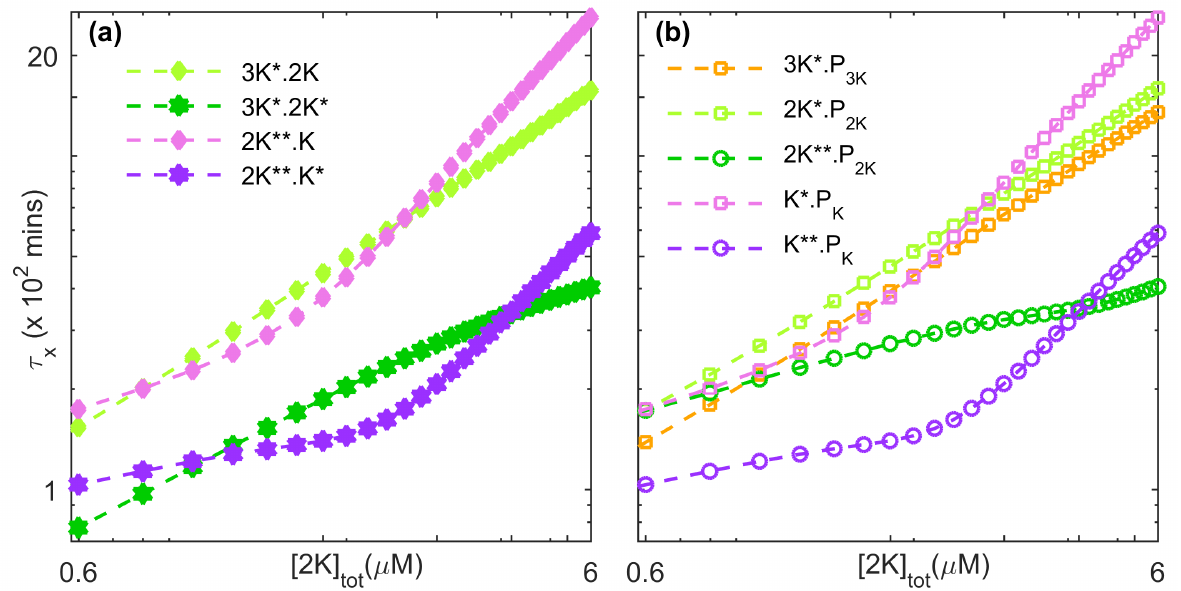}
\end{center}
\caption{Protein complexes in the MAPK cascade exhibit relaxation
behavior occurring over a broad range of time-scales.
Decay of activity is shown after withdrawing an applied stimulus of amplitude $S = 1.2 \times 10^{-6} \mu M$.
The relaxation times $\tau_{\rm x}$ of the different molecular
species, viz., (a) the protein complexes between non-phosphorylated
and singly phosphorylated (non-active) kinase proteins and the
doubly phosphorylated (active) kinase protein of the preceding layer,
and (b) the protein complexes between the phosphorylated (singly- or
doubly-) kinase proteins and the phosphatase that carries out
dephosphorylation in the corresponding layer of the MAPK cascade, vary
with the total concentration of MAP2K. The nature of this dependence
is distinct for lower and higher values of $[2K]_{tot}$. For both
panels,
$[K]_{tot} = 0.8 \mu M$ and $[3K]_{tot} = 0.0020 \mu M$. The total
concentrations of the phosphatases are
provided in Table~\ref{table:phosphatases}.
}
\label{fig:fig13}
\end{figure*}

\begin{figure*}
\begin{center}
\includegraphics[ ]{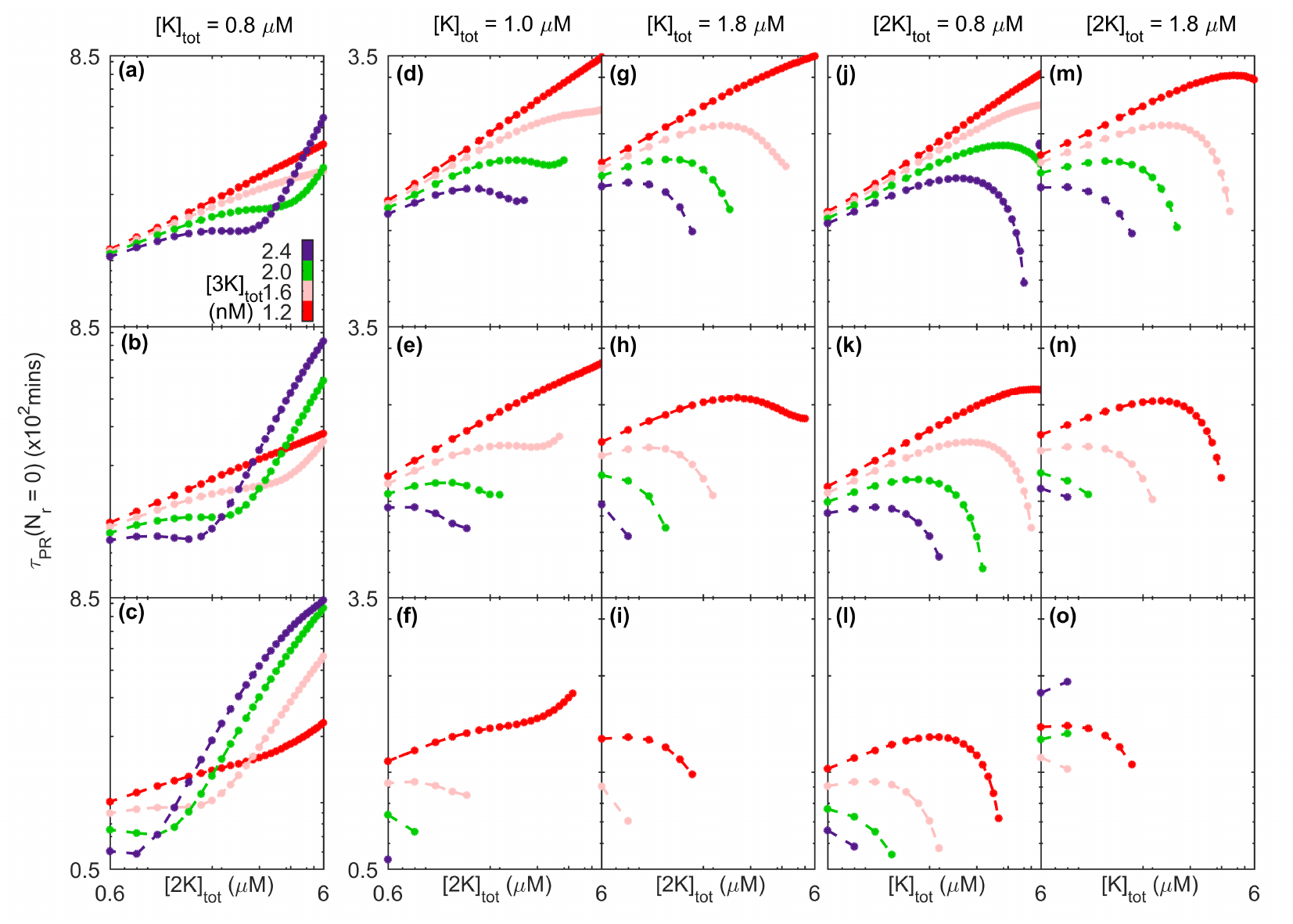}
\end{center}
\caption{Dependence of the primary recovery time $\tau_{PR}$ on (a-i) the total concentration of MAP2K ($[2K]_{tot}$) and on (j-o) the total concentration of MAPK ($[K]_{tot}$) for different values of the
total concentration of MAP3K ($[3K]_{tot}$), obtained upon removing
stimuli having different amplitudes $S$. Panels (a,d,g,j,m) are for
$S=0.8 \times 10^{-6} \mu M$, panels 
(b,e,h,k,n) are for $S=1.2 \times 10^{-6} \mu M$, and panels
(c,f,i,l,o) are for $S=2.0
\times 10^{-6} \mu M$. We have only considered situations where the
system reaches a steady state upon application of a time-invariant
stimulus, and that do not show any reverberatory activity ($N_{r} =
0$) during relaxation to the resting state.
The curves in panels (a-i) are obtained for
different values of $[K]_{tot}$, namely, (a-c) $[K]_{tot} = 0.8 \mu M$,
(d-f) $[K]_{tot} = 1.0 \mu M$, and (g-i) $[K]_{tot} = 1.8 \mu M$.
The curves in panels (j-o) are obtained for different values of
$[2K]_{tot}$, namely,
(j-l) $[2K]_{tot} = 0.8 \mu M$, and (m-o) $[2K]_{tot} = 1.8 \mu M$.
The total concentrations of the phosphatases for all panels are given in
Table~\ref{table:phosphatases}.
}
\label{fig:fig14}
\end{figure*}

\begin{figure*} 
\begin{center}
\includegraphics[ ]{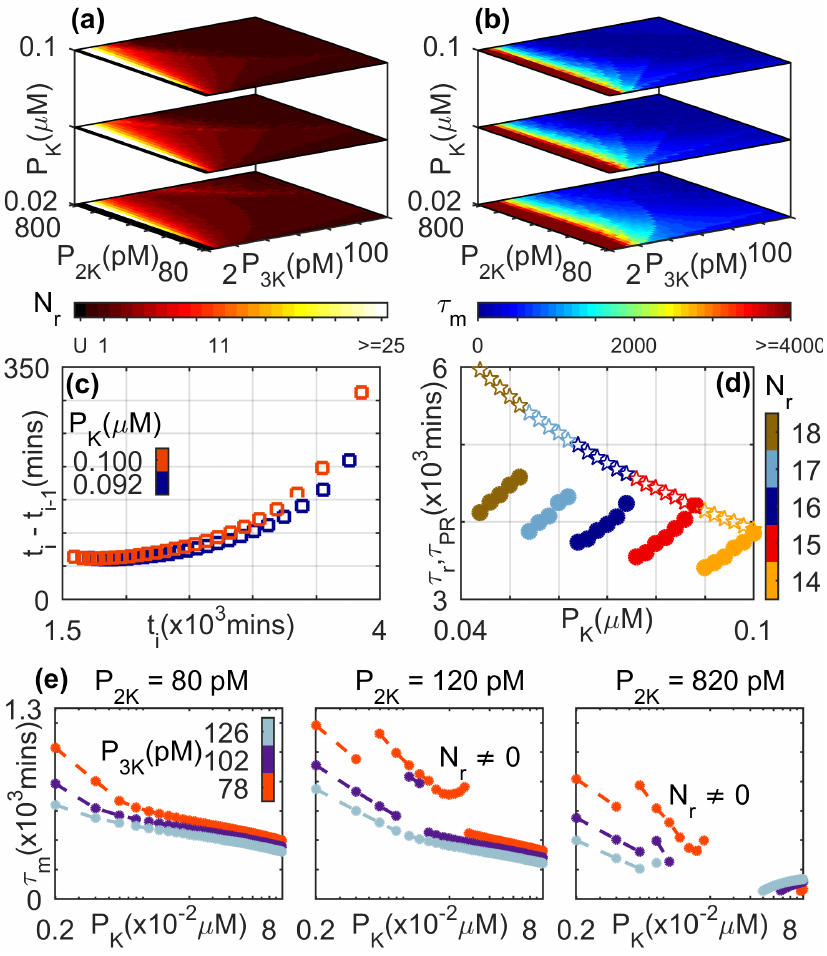}
\end{center}
\caption{Dependence of reverberatory activity on the total
concentrations of the phosphatases MAPK P'ase ($[P_{K}]$),
MAP2K P'ase, ($[P_{2K}]$) and MAP3K P'ase ($[P_{3K}]$). (a) The number
of spikes $N_{r}$ and (b) the total memory time $\tau_m$
(in minutes) observed on withdrawing an applied stimulus of amplitude
$S=2.0 \times 10^{-6} \mu M$. Situations where the primary recovery
time is longer than a maximum or cut-off value (see Methods), such
that the reverberatory nature of the dynamics cannot be properly
measured, are
indicated by the color corresponding to ``U''.
(c) The interval between successive spikes $i-1$ and $i$ increases
with time ($t_{i}$ being the time of occurrence of the $i$th
spike).
As the MAPK P'ase concentration is increased, the durations of these intervals are seen to increase. The total concentrations of the other
two phosphatases are maintained at $[P_{2K}] = 680 pM$ and $[P_{3K}] = 10 pM$.
(d) The variation of primary recovery time $\tau_{\rm PR}$
(stars) and the total duration of reverberatory activity $\tau_{r}$
(filled circles) as a function of total MAPK P'ase concentration are shown
for different values of $N_r$ (indicated by the color bar). 
While $\tau_{\rm PR}$ decreases monotonically with increasing
$[P_{K}]$, $\tau_{r}$ shows a more complex dependence ($[P_{2K}] = 200
pM$ and $[P_{3K}] = 6 pM$). 
(e) Dependence of the total memory time $\tau_{m}$ on total MAPK P'ase
concentration ($[P_K]$ shown in log scale) for
different total concentrations of MAP2K P'ase (values indicated above
each of the three panels) and MAP3K P'ase
(indicated using different colors as shown in the color bar). 
Note that we consider only situations where the system attains a
steady state on
maintaining stimulation. For details of the total concentrations of the kinases, see
Table~\ref{table:kinases47}.
}

\label{fig:fig15}
\end{figure*}

\begin{figure*} 
\begin{center}
\includegraphics[ ]{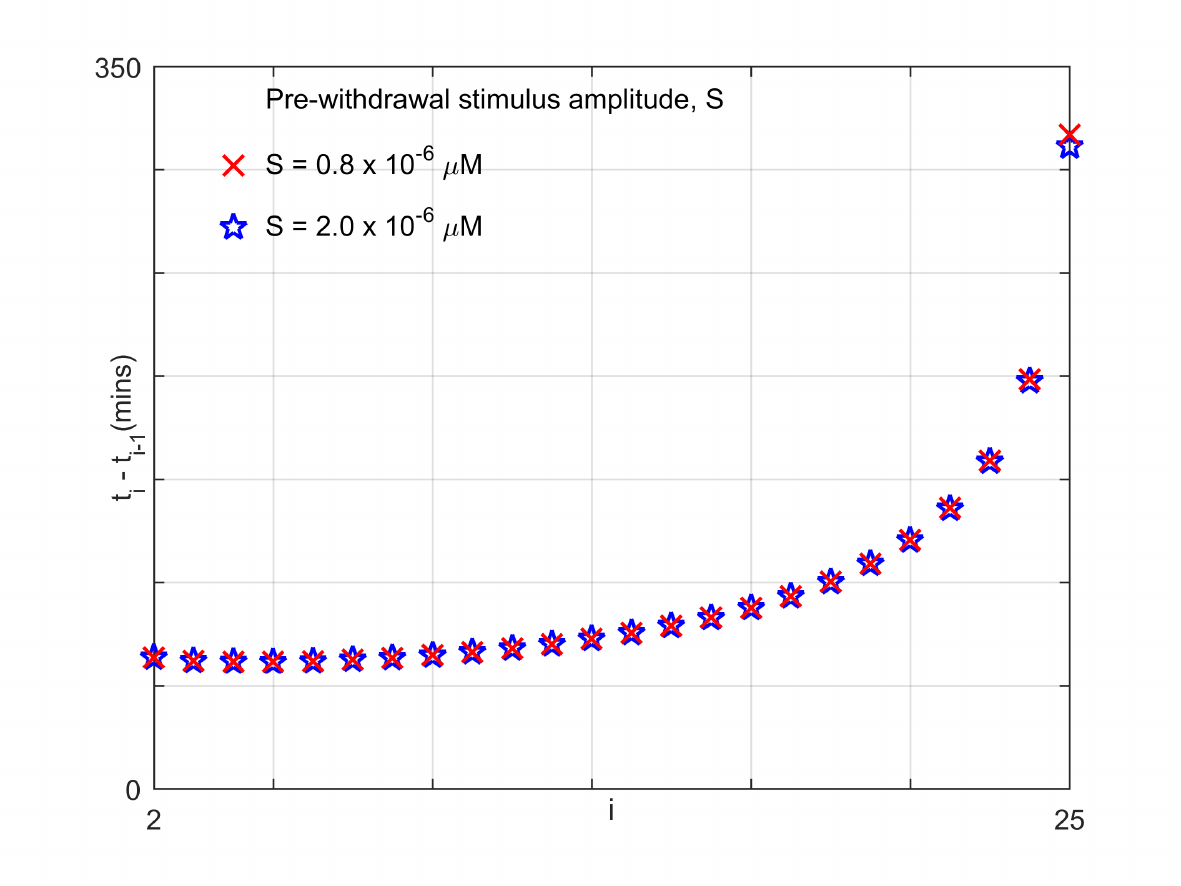}
\end{center}
\caption{The time interval between successive spikes $i-1$ and $i$
obtained after removing a stimulus, increases
with the number of spike events ($i$ being the event number of the
$i$th spike). The trend appears to be independent of the
stimulus amplitude $S$. The total concentrations of the phosphatases
are $P_{K} = 0.1 \mu M$, $P_{2K} = 680 pM$ and $P_{3K} = 10 pM$,
respectively. The total concentrations of the kinases are provided in
Table~\ref{table:kinases47}.
}
\label{fig:fig16}
\end{figure*}

\end{document}